\documentclass[reprint, nofootinbib, 
amsmath,amssymb, aps, floatfix, 
twoside]{revtex4-1}

\usepackage[utf8]{inputenc}
\usepackage[brazilian]{babel}
\usepackage[T1]{fontenc}
\usepackage{array}
\usepackage{amsmath}
\usepackage{amsfonts}
\usepackage{amssymb}
\usepackage{amstext}
\usepackage{amsthm}

\usepackage{color}
\usepackage{listings}
\usepackage{placeins}
\usepackage{graphicx}
\usepackage[colorlinks=true,citecolor=blue,linkcolor=blue,urlcolor=blue]{hyperref}
\usepackage{mathtools}
\usepackage{times}
\usepackage{latexsym}
\usepackage[running,mathlines]{lineno}
\usepackage{verbatim}
\usepackage{bm}
\usepackage{ulem}
\usepackage{tcolorbox}

\usepackage{listings}
\usepackage{xcolor}  
\usepackage{ragged2e}

\definecolor{codegreen}{rgb}{0,0.6,0}
\definecolor{codegray}{rgb}{0.5,0.5,0.5}
\definecolor{codepurple}{rgb}{0.58,0,0.82}
\definecolor{backcolour}{rgb}{0.95,0.95,0.92}

\tcbuselibrary{minted,breakable,skins,xparse}
\definecolor{bg}{gray}{0.95}

\definecolor{codegreen}{rgb}{0,0.6,0}
\definecolor{codegray}{rgb}{0.5,0.5,0.5}
\definecolor{codepurple}{rgb}{0.58,0,0.82}
\definecolor{backcolour}{rgb}{0.95,0.95,0.92}

\lstdefinestyle{mystyle}{
    backgroundcolor=\color{backcolour},   
    commentstyle=\color{codegreen},
    keywordstyle=\color{magenta},
    numberstyle=\tiny\color{codegray},
    stringstyle=\color{codepurple},
    basicstyle=\ttfamily\footnotesize,
    breakatwhitespace=false,         
    breaklines=true,                 
    captionpos=b,                    
    keepspaces=true,                 
    numbers=left,                    
    numbersep=5pt,                  
    showspaces=false,                
    showstringspaces=false,
    showtabs=false,                  
    tabsize=2
}

\newtheoremstyle{mytheoremstyle} 
  {0.5cm}                        
  {0.5cm}                        
  {}                     
  {}                             
  {\bfseries}                    
  {.}                            
  {0.5em}                        
  {}                             
\theoremstyle{mytheoremstyle}
\newtheorem{teorema}{Teorema}[section] 
\theoremstyle{mytheoremstyle}
\theoremstyle{mytheoremstyle}

\usepackage{hyperref}

\begin{document}
\raggedbottom

\title{\Large{ Introdução às redes neurais para Físicos} \\
(An introduction to Neural Networks for Physicists)}

\author{Gubio G. de Lima$^\dagger $}
\affiliation{Universidade Federal de São Carlos, Departamento de Física, São Carlos, SP, Brasil}

\author{Gustavo Café de Miranda$^\dagger $} \email[Email:]{ gcaf0125@uni.sydney.edu.au}
\affiliation{University of Sydney, Sydney, NSW, Austrália}

\author{Tiago de S. Farias }
\affiliation{Universidade Federal de São Carlos, Departamento de Física, São Carlos, SP, Brasil}

\address{$^\dagger $ Ambos os autores contribuíram igualmente para este trabalho.}

\begin{abstract} 

As técnicas de aprendizado de máquina emergiram no  contexto científico e se desenvolveram como ferramentas poderosas para enfrentar uma ampla gama de desafios na sociedade. A integração dessas técnicas com a física tem conduzido a abordagens inovadoras na compreensão, controle e simulação de fenômenos físicos. Este artigo visa proporcionar uma introdução prática às redes neurais e seus conceitos fundamentais, destacando perspectivas recentes dos avanços na interseção entre modelos de aprendizado de máquina e sistemas físicos. Além disso, apresentamos um material prático para orientar o leitor em seus primeiros passos na aplicação de redes neurais para resolver problemas físicos. Como exemplo ilustrativo, fornecemos quatro aplicações de complexidades crescentes para o problema de um pêndulo simples, a saber: \textit{fit} de parâmetros da Equação Diferencial Ordinária (EDO) do pêndulo para aproximação de ângulo pequeno; \textit{Physics Informed Neural Networks} (PINNs) para encontrar soluções da EDO do pêndulo em ângulo pequeno; \textit{Autoencoders} em conjunto de dados de imagens do pêndulo para estimação de dimensionalidade do espaço de parâmetros do problema físico; uso de arquiteturas \textit{Sparse Identification of Non-Linear Dynamics} (SINDy) para descoberta de modelos e expressões analíticas para o problema do pêndulo não linear (ângulos grandes). 

\begin{description}
\item[Palavras-chave] Física clássica; Redes neurais; Tutorial.
\end{description}

\bigbreak 

Machine learning techniques have emerged in the scientific context and have developed into powerful tools for addressing a wide range of challenges in society. The integration of machine learning methods with physics has led to innovative approaches in understanding, controlling, and simulating physical phenomena. This article aims to provide a practical introduction to neural network and their basic concepts. It presents some perspectives on recent advances at the intersection of machine learning models with physical systems. We introduce practical material to guide the reader in taking their first steps in applying neural networks to Physics problems. As an illustrative example, we provide four applications of increasing complexity for the problem of a simple pendulum, namely: parameter fitting of the pendulum's ODE for the small-angle approximation; application of Physics-Inspired Neural Networks (PINNs) to find solutions of the pendulum's ODE in the small-angle regime; \textit{Autoencoders} applied to an image dataset of the pendulum's oscillations for estimating the dimensionality of the parameter space in this physical system; and the use of Sparse Identification of Non-Linear Dynamics (SINDy) architectures for model discovery and analytical expressions for the nonlinear pendulum problem (large angles).

\begin{description}
\item[Keywords] Classical Physics; Neural network; Tutorial.
\end{description}
\end{abstract}

\maketitle
\section{Introdução}

O aprendizado de máquina, em inglês \textit{Machine Learning} (ML), tem ganhado muita atenção nos últimos anos devido ao seu sucesso em tarefas comerciais, industriais e especialmente no setor de serviços \cite{ige2025machine,SAHUT2025108684,KHEDR2024100379,geron, NathanKutz2017}. Hoje, os algoritmos baseados em ML, comumente associados ao termo Inteligência Artificial (IA), estão profundamente integrados às tecnologias digitais. Essas técnicas são notoriamente bem-sucedidas no tratamento de grandes volumes de dados, permitindo  a solução de problemas complexos mediante métodos estatísticos \cite{carleo_machine_2019}. Esse sucesso tem atraído a atenção de pesquisadores, destacando-se como uma poderosa ferramenta estatística e uma potencial aliada na exploração científica.

Embora promissor, o uso de ML pode fornecer modelos que são, por vezes, opacos, frequentemente tratados como ``caixas-pretas'', onde não é possível determinar as relações causais que conduziram o algoritmo aos seus resultados. Por esta razão, estes algoritmos podem ser recebidos com ceticismo por membros da comunidade científica, que têm preferência por modelos mais interpretáveis e com fundamentos teóricos claros \cite{carleo_machine_2019}. Contudo, tal ceticismo não foi impeditivo para a investigação de novas aplicações e algoritmos em diferentes setores no meio científico, especialmente em áreas inclinadas às aplicações, posicionando o ML como uma ferramenta promissora para o avanço da ciência. Nos parágrafos subsequentes, apresentaremos alguns dos casos em que esses esforços demonstraram êxito no meio científico, com enfoque especial na área de física.

Embora a área de ML tenha ganhado destaque significativo nos últimos anos, como exemplo o Prêmio Nobel de Física em 2024, muitos dos modelos utilizados atualmente vêm sendo desenvolvidos e aprimorados há várias décadas \cite{LiGilbert2024}. A partir da segunda metade do século XX, com os avanços na capacidade computacional, equipados com técnicas estatísticas e métodos matemáticos, estabeleceram-se os fundamentos do campo do ML. Nesse contexto, diferentes abordagens começaram a emergir, incluindo a investigação do comportamento dos neurônios biológicos, que inspirou a criação de algoritmos de redes neurais artificiais (em inglês, \textit{Artificial Neural Networks)} com o objetivo de replicar, de forma simplificada, os processos de aprendizado observados em organismos vivos~\cite{Schmidhuber2015}. 

As redes neurais artificiais dependem de abordagens de treinamento que orientam como os modelos interagem com os dados para resolver problemas específicos. Esses diferentes paradigmas de treinamento, como o aprendizado supervisionado, o aprendizado reforçado e o aprendizado não supervisionado, definem maneiras distintas de explorar a relação entre os dados e o modelo, permitindo sua aplicação em uma ampla variedade de contextos.

O paradigma de ML supervisionado envolve o treinamento de modelos  que utilizam conjuntos de dados rotulados, nos quais cada entrada está associada a um rótulo ou valores desejados (saída). Esse processo permite que os algoritmos tentem aproximar um mapa que represente a relação entre os dados de entrada e saídas~\cite{geron}. Algoritmos comuns de aprendizado supervisionado em ML são\footnote{Em inglês: \textit{K-Nearest Neighbors}, \textit{Logistic Regression}, \textit{Support Vector Machines} (SVM), \textit{Decision Trees}, and \textit{Artificial Neural Networks}.}: K-vizinhos mais próximos \cite{KNN,KNN2}, Regressão Logística \cite{logiste-regression1,logiste-regression2}, Máquinas de Vetores de Suporte \cite{SVM1,SVM2}, Árvores de Decisão \cite{decision-tree1,decision-tree2}, e redes neurais artificiais \cite{supervised-nn}. 

Por outro lado, o aprendizado não supervisionado é caracterizado pelo uso de algoritmos que lidam com dados não rotulados, i.e.,  há uma estrutura subjacente nos dados que não é explicitamente conhecida, mas cujos padrões deseja-se identificar. Nesse contexto, os modelos devem identificar  relações e características exclusivamente a partir dos dados de entrada, utilizando funções apropriadas para otimizar um objetivo que visa identificar as relações inerentes entre os dados fornecidos. Otimizar estas funções leva a uma saída que deve ser consistente com restrições (às vezes adicionais) definidas pelo treinamento. Historicamente, o aprendizado não supervisionado se tornou particularmente atrativo com a publicação de artigos de redes neurais inspiradas em modelos biofísicos do córtex visual em felinos \cite{Schmidhuber2015}.
Alguns algoritmos amplamente discutidos na literatura para esse tipo de aprendizado incluem: K-Médias \cite{unsupervised},  Density-Based Spatial Clustering of Applications with Noise (DBSCAN) \cite{unsupervised},  Principal Component Analysis (PCA) \cite{unsupervised}, \textit{t-Distributed Stochastic Neighbor Embedding} (t-SNE) \cite{supervised-nn}, e \textit{autoencoders} \cite{Marquardt2021,Goodfellow-et-al-2016,SINDyAutoencoder_steven2019}. 

No aprendizado por reforço, o modelo, frequentemente representado como um agente, aprende a tomar decisões através da interação contínua com um ambiente dinâmico, que pode ser total ou parcialmente observável, baseado em um processo de tentativa e erro. Nesse contexto, o agente executa uma série de ações em resposta a estados observados do ambiente e, para cada ação tomada, ele recebe uma recompensa ou penalidade, dependendo de quão benéfica ou prejudicial a ação foi para atingir um objetivo definido. O objetivo central do agente é aprender uma estratégia que define a melhor ação a ser tomada em cada estado possível do ambiente, visando maximizar a recompensa acumulada ao longo do tempo. O aprendizado por reforço demonstra eficácia em uma ampla variedade de aplicações complexas, como jogos, robótica, sistemas autônomos de controle e recomendação de produtos, e outros sistemas cuja solução pode ser escrita em termos de agentes e estratégias. Exemplos notáveis incluem o treinamento de agentes que superam humanos em jogos de tabuleiro e videogames, como o AlphaGo \cite{alphago}, Atari \cite{atari} e o desenvolvimento de sistemas de controle em robôs que aprendem a se movimentar e a manipular objetos em ambientes reais \cite{robothand}.

Historicamente, no campo da física, a pesquisa em física de partículas foi pioneira na adoção de técnicas de ML, em parte devido à necessidade de lidar com grandes volumes de dados em tempo real. As primeiras aplicações de ML nesse campo focaram em sistemas com grandes bancos de dados, que precisavam ser processados rapidamente. Na física de altas energias, por exemplo, o uso de árvores de decisão em experimentos com colisores de partículas ganhou destaque. Nesses experimentos, a rápida ativação dos detectores exige métodos computacionais eficientes para a classificação e armazenamento de dados enquanto os eventos ocorrem \cite{Gligorov2013}. Posteriormente, para a mesma tarefa, o uso de redes neurais mostrou-se superior, superando os métodos tradicionais de análise de dados \cite{baldi2014}.

Outro exemplo de aplicação de ML na física de partículas envolve a investigação de modelos que dependem de expansões perturbativas da Teoria Quântica de Campos, modelos de interação partícula-detector e modelos fenomenológicos, todos focados em descrever interações em diferentes escalas de comprimento. Esses modelos geralmente apresentam parâmetros livres que precisam ser ajustados com base em dados experimentais. O uso de redes neurais para ajustar esses parâmetros, conectando modelos teóricos a resultados experimentais, tornou-se uma aplicação comum tanto em física de partículas quanto em cosmologia \cite{carleo_machine_2019,Feickert}.

Como mostrado pelos exemplos anteriores, as redes neurais têm demonstrado sua eficácia em tarefas computacionalmente complexas, com a etapa de otimização representando uma parte significativa do esforço computacional. No entanto, uma vez otimizadas, essas redes podem ser reutilizadas em seu domínio de aplicação com novos dados, um processo que geralmente exige menor intensidade computacional.

Estes resultados têm se mostrado promissores na literatura em física de muitos corpos, tanto clássicos como quânticos, onde soluções para caracterização de fases em modelos de spins (i.e. \textit{Ising} 2D)  são uma questão central \cite{carrasquila2017, VonNieuwenburg2017}. Em matéria condensada e estudo de materiais, ML é usada, aliado a outros métodos, para cálculo de energias em \textit{Density Functional Theory} (DFT), sendo sensivelmente mais rápida que métodos tradicionais \cite{Smith2017}. 

Uma segunda abordagem, de especial interesse para a comunidade científica, por fornecer um meio-termo entre clareza e opacidade de modelos, 
foi pensada  para agregar informação física nos modelos, integrando leis ou restrições físicas em modelos de ML, apresentando novas oportunidades para as pesquisas científicas.

Um dos principais métodos nessa abordagem é conhecido como  \textit{Physics-Informed Machine Learning} (PIML)\cite{PIML-review,PIML-review-nature,PIML-review2} e \textit{Physics-Informed Neural Network} (PINN)\cite{PINN-review,PINN-review2}, cujo uso pode ser encontrado em publicações de diversas áreas de pesquisa na Física, como: mecânica dos fluidos \cite{PINN-review-fluid,PINN-review-fluid2}; quantificação da incerteza \cite{Psaros}; sistemas dinâmicos \cite{WangandR}; sistema quântico de muitos corpos \cite{Carrasquilla}, sistemas fotônicos\cite{nn-phonics}; óptica clássica \cite{nn-optics}, entre outros \cite{outro1,outros2,outros3,outros4}. Existem outros trabalhos com paradigmas de modelagem híbrida que integram ML com conhecimento físico \cite{Pinn-hibrido0,Pinn-hibrido1,Pinn-hibrido2,Fouriernn,OML,OML2,IAfeymann,PhyCV,PICV}.
Além de adicionar informações físicas ao modelo, há na literatura aplicações que invertem o problema de pesquisa, ou seja, utilizam-se de ML para identificar características físicas. Exemplos incluem determinar equações que descrevem o sistema por meio de regressão simbólica \cite{MaxTegmark2020} ou o uso de autodiferenciação em bibliotecas avançadas de ML \cite{Baydin-automaticdiff}.

A literatura científica na interseção entre física e ML tende a ser segmentada em nichos que exigem conhecimentos avançados, muitas vezes ao nível de pós-graduação. Pensando em tornar esse conhecimento mais acessível, aqui apresentamos materiais voltados para um público mais amplo, como graduandos em física que já completaram o ciclo básico, bem como estudantes de engenharia. Para isso, utilizaremos problemas de mecânica clássica combinados com técnicas de redes neurais, de modo a facilitar a compreensão e o aprendizado desses temas complexos.

Neste estudo, discutimos os conceitos fundamentais das redes neurais, começando com o Perceptron no Capítulo \ref{sec:perceptron}, onde apresentamos um passo-a-passo de como construir e aplicá-lo.
Em seguida, abordamos as redes neurais profundas no Capítulo \ref{sec:RNProfunda}, ampliando alguns conceitos, introduzindo novos e apresentando as arquiteturas mais comuns.
No Capítulo \ref{sec:Capitulo3}, apresentaremos quatro aplicações distintas de redes neurais no  modelo unidimensional do pêndulo simples. 
Na primeira abordagem (\ref{sec:EX:Pendulo}), foi realizado o aprendizado supervisionado com uma rede neural para determinar qual a constante do sistema físico (constante gravitacional). Em seguida, exploramos o uso de redes neurais profundas e autodiferenciação para resolver a equação diferencial associada ao pêndulo (\ref{sec:EX-EDO}). Posteriormente, investigamos a utilização de \textit{autoencoders} (\ref{sec:EX:AutoencoderPendulo}) para inferir o espaço latente do sistema e estimar sua dimensão, além de introduzir o uso destes modelos para filtragem de ruídos em dados e imagens. Por fim, na seção (\ref{sec:EX:SINDY}) usaremos \textit{autoencoders} SINDy para descobrir a equação diferencial (e não sua solução) do sistema do pêndulo não linear (para ângulos grandes).

\section{Aspectos básicos de Redes neurais }
\label{sec:capitulo2}

Ao longo das últimas décadas, as redes neurais artificiais foram investigadas com o intuito de modelar o comportamento dos neurônios biológicos. A associação entre os fenômenos biológicos e a inteligência motivou a pesquisa além do campo da neurofisiologia, originando, em 1943, o modelo proposto por McCulloch e Pitts, considerado o primeiro estudo a descrever formalmente uma rede neural artificial \cite{Schmidhuber2015}. Embora a terminologia específica tenha sido cunhada apenas mais tarde, esse trabalho estabeleceu as bases conceituais para os modelos computacionais de aprendizagem. Em 1958, Rosenblatt propôs o perceptron, um modelo teórico inspirado em princípios do funcionamento do cérebro, que buscava explicar como organismos poderiam aprender, reconhecer padrões e generalizar informações de forma probabilística \cite{rosenblatt1958}. Nas décadas seguintes, novas arquiteturas e métodos de aprendizagem foram desenvolvidos, consolidando as redes neurais como um dos pilares da inteligência artificial moderna \cite{Schmidhuber2015}.

A década de 2010 foi particularmente prolífica para o avanço do aprendizado de máquina e das redes neurais. Um dos fatores decisivos foi a popularização das Unidades de Processamento Gráfico (em inglês, \textit{Graphics Processing Units}, GPUs), originalmente projetadas para executar milhares de operações matemáticas em paralelo. Diferentemente das CPUs, que possuem um número limitado de núcleos otimizados para tarefas sequenciais, as GPUs contam com milhares de núcleos mais simples, especializados no processamento simultâneo de grandes volumes de dados. Essa capacidade de paralelismo massivo é ideal para o treinamento de redes neurais, que envolve álgebra linear em larga escala, reduzindo drasticamente o tempo necessário para ajustar os parâmetros dos modelos. Como resultado, as redes neurais começaram a superar os algoritmos determinísticos tradicionais em diversas áreas de grande prestígio.

No campo do processamento de imagens, por exemplo, redes neurais convolucionais superaram os métodos clássicos no desafio ImageNet, que antes eram dominados por técnicas como SIFT (Scale-Invariant Feature Transform) combinadas com Máquinas de Vetores de Suporte (SVM) \cite{Schmidhuber2015, Russakowsky2015}. No xadrez, a abordagem baseada em busca heurística e poda alfa-beta, como a do motor Stockfish \textit{Stockfish}, foi superada pela AlphaZero, uma rede neural que aprendeu o jogo do zero \cite{Alphazero}. O sucesso se estendeu a outros domínios com ampla cobertura midiática, como a vitória do AlphaGo no jogo de Go \footnote{deepmind.google/technologies/alphago/} e o surgimento de modelos de linguagem avançados como o ChatGPT \cite{Ray2023ChatGPTAC}.

\subsection{Perceptron.}
\label{sec:perceptron}

\begin{figure}[!h]
    \centering
    \includegraphics[width=1.0\linewidth]{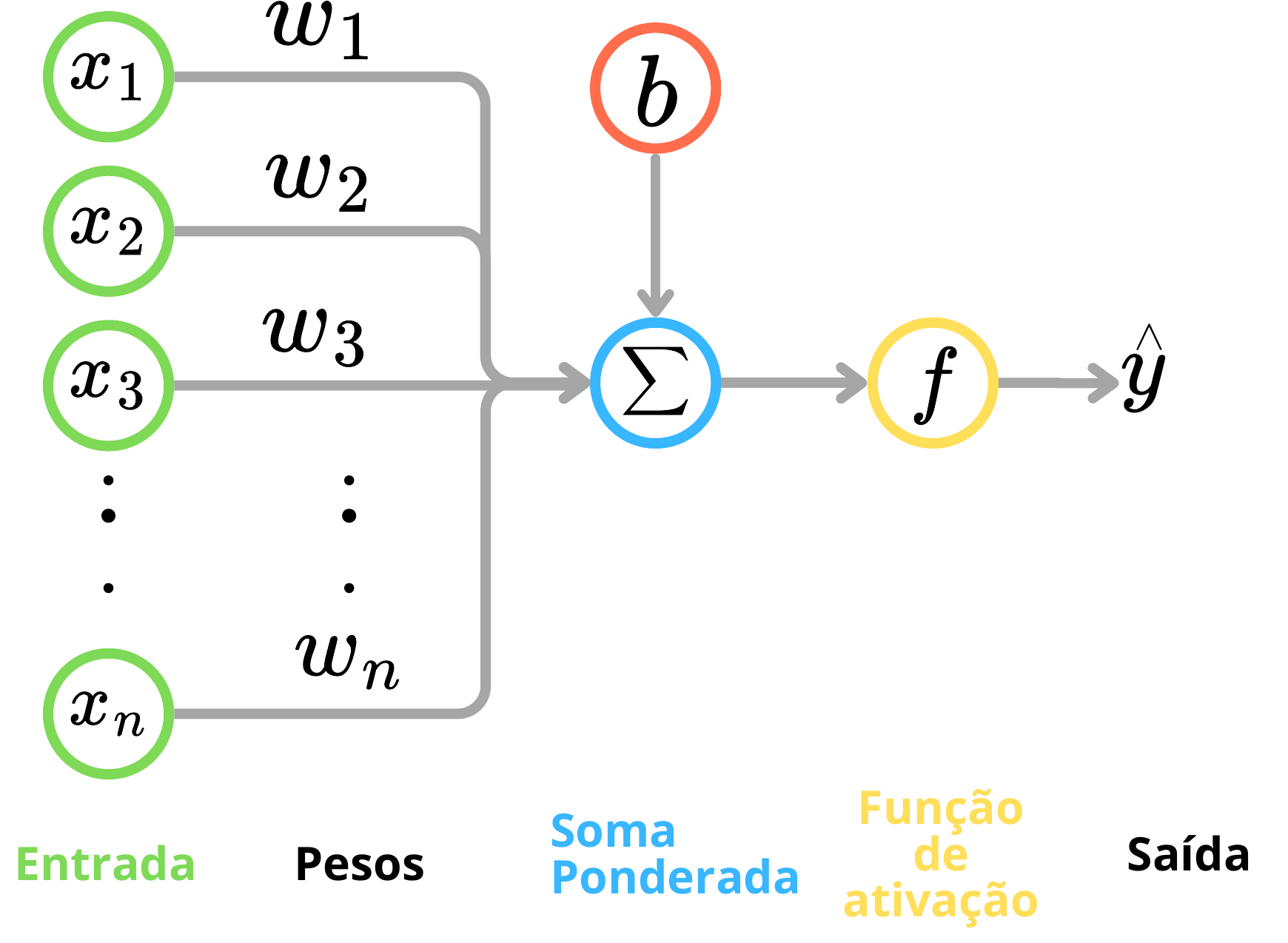}
    \caption{Ilustração das etapas de um Perceptron: $x_{ij}$ representa os elementos do vetor de entrada $\mathbf{x}_j$ (em verde) associado à amostra $j$,  estes elementos são multiplicados um a um por pesos $w_i$ e somados a um \textit{bias} $b$. O resultado desta soma (em azul) serve de argumento para uma função de ativação $f$ (em amarelo), produzindo a saída $\hat{y}$.}
    \label{fig:perceptron}
\end{figure}

Nos algoritmos de redes neurais artificiais, o neurônio é definido como a unidade fundamental responsável pelo processamento de informação. Inspirados pelos neurônios biológicos, os neurônios artificiais recebem um conjunto de entradas, processam essas informações aplicando um conjunto de operações matemáticas que produzem uma saída. Um dos modelos mais simples de neurônios artificiais é o Perceptron \cite{rosenblatt1958, Marquardt2021}. As redes neurais são tipicamente representadas como grafos direcionados, onde os neurônios correspondem aos vértices e as conexões entre eles, denominadas sinapses no contexto biológico e pesos no contexto algorítmico, são representadas por arestas.  O Perceptron pode ser representado conforme ilustrado na Figura \ref{fig:perceptron}. 

De maneira simplificada, no modelo biológico, interpretamos as entradas como mudanças na diferença de potencial eletrostático na vizinhança $x_i$ do neurônio. Já os pesos $w_i$ representam a intensidade ou sensibilidade da conexão com cada parte $i \in \{1,\ldots,n\}$ da vizinhança.
O \textit{bias} $b$ pode ser entendido como um campo médio que influencia o campo efetivo no neurônio ou como um limiar de ativação que o neurônio deve superar para ser ativado. 
Esses elementos atuam em conjunto sendo somados na operação $\Sigma$, cuja saída serve como argumento para a função de ativação $f$. Esta função $f$ determina se o neurônio é ativado (isto é, ``dispara'' um sinal para outros neurônios) ou permanece inativo.  As analogias com a biologia, no entanto, apresentam limitações, pois o modelo matemático não consegue reproduzir integralmente a complexidade dos sistemas biológicos. Por exemplo, o sinal neuronal possui uma dependência temporal intrínseca, algo que não é considerado no perceptron clássico, cuja dinâmica é essencialmente estática. Para incorporar essa característica temporal, foram propostas arquiteturas mais sofisticadas, como as redes neurais recorrentes, que introduzem memória ao sistema ao levar em conta o estado anterior da rede \cite{Lipton2015}. 
Neste trabalho essa analogia  tem um caráter meramente didático para os propósitos deste artigo e, portanto, não utilizaremos esses termos biológicos nas seções subsequentes. No contexto de redes neurais artificiais, a saída $\hat{y}$ do Perceptron indica a resposta do neurônio a um dado conjunto de entradas, refletindo se o estímulo foi suficiente para ativar o neurônio.

A linguagem \textit{Python} se estabeleceu como o padrão para aprendizado de máquina devido à sua sintaxe simples e legível, que permite ao programador focar nos conceitos algorítmicos em vez de se prender a complexidades sintáticas. Mais importante ainda, \textit{Python} é sustentado por um ecossistema de bibliotecas robusto e maduro, como \textit{NumPy}, para computação numérica, e frameworks especializados como \textit{Scikit-learn}, \textit{TensorFlow} e \textit{PyTorch}, que simplificam drasticamente o desenvolvimento e o treinamento de modelos de ML. Por essa razão, para facilitar a compreensão dos leitores, detalharemos a seguir cada etapa do funcionamento do Perceptron, abordando tanto os aspectos matemáticos envolvidos quanto sua implementação prática nesta linguagem de programação.

\subsubsection{Exemplo didático: Classificação binária}

Para compreender os conceitos representados na Figura \ref{fig:perceptron}, partiremos de um exemplo de aprendizado supervisionado, que será detalhado nos parágrafos a seguir. Seguindo a ordem de execução de um Perceptron, da esquerda para a direita. São eles: as entradas (dados), os pesos, a soma ponderada com a função de ativação e, por fim, a saída da rede neural.

\textbf{ Entradas (\textit{inputs}):} A primeira etapa envolve os dados de entrada, representados pelos círculos verdes à esquerda na Figura \ref{fig:perceptron}. Cada elemento $x_i$ representa uma variável independente. Estas entradas são valores numéricos que caracterizam as variáveis do problema em análise.
Por exemplo, podemos ter uma tabela ou planilha com uma lista de produtos e suas características, ou um banco de dados com informações sobre animais, plantas ou pessoas, etc. As entradas também podem ser informações de imagens \cite{imagem-nn}, sons \cite{sound-nn} ou textos \cite{text-nn}, ou até mesmo os valores de saída de outros neurônios.

\begin{figure}[!ht]
    \centering
    \includegraphics[width=1\linewidth]{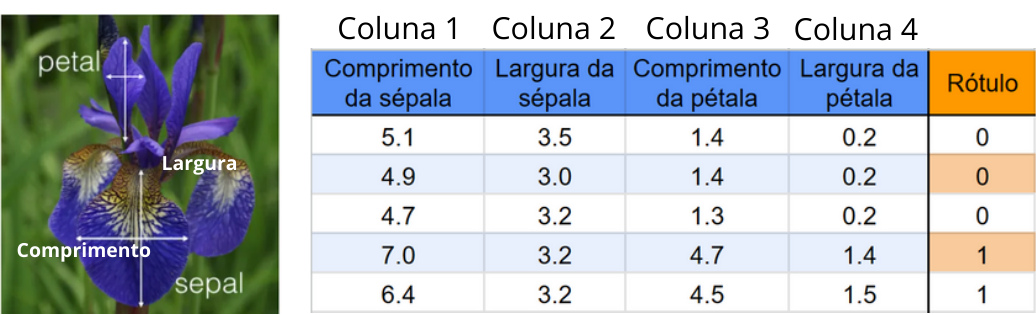}
    \caption{Banco de dados de Iris (\textit{Iris dataset}), onde estão explicitas as primeiras 5 linhas do \textit{data-set}, com informações sobre flores, como comprimento e largura da sépala ou da pétala.}
    \label{fig:iris}
\end{figure}

Para exemplificar, selecionamos um conjunto de dados resultante da catalogação de flores, que contém características do comprimento e largura das pétalas e sépalas. Este banco de dados é conhecido como \textit{Iris dataset}~\cite{scikit-learn_iris}, comumente utilizado para estudar problemas de classificação. Na Figura \ref{fig:iris}, temos as primeiras 5 linhas do banco de dados, onde cada linha representa uma flor catalogada e cada coluna uma característica. Cada linha é um vetor $\mathbf{x_j}$ (com $j$ representando a linha específica do conjunto de dados), dado por \footnote{Adotamos o padrão americano de pontuação decimal (ponto como separador decimal), nos códigos e equações.} :
\begin{equation}
    \mathbf{x}_j = [ x_{j1} , x_{j2}, x_{j3}, \ldots, x_{jn} ], 
    \label{eq-1}
\end{equation}
\noindent cada elemento $x_{ji}$ do vetor, varia de $i \in \{1, \cdots, N\}$  onde $N$ é o número de colunas do banco de dados e $j \in \{0, \cdots, M\}$  onde $M$ é o número de dados (ou amostragem). A  última coluna, com $i =N+1$, é reservada aos rótulos, que neste caso representam a classificação das espécies e a coluna para $i =0$ tem valor específico, que será mencionado no próximo passo. 

Para o caso do exemplo ilustrado na Figura \ref{fig:iris}, temos 4 características (colunas) para a entrada e uma característica para o rótulo. Então, a representação matemática de uma linha específica ($j=0$) da Figura \ref{fig:iris} pode ser expressa por:

\begin{align*}
    \mathbf{x}_0 &= [ x_{01} , x_{02} , x_{03} ,  x_{04}],
\end{align*}
\begin{equation}
    \mathbf{x}_0 = [5.1, 3.5, 1.4, 0.2].
    \label{equationX0}
\end{equation}

Para acessar o conjunto de dados do \textit{Iris dataset}, podemos utilizar a biblioteca \textit{scikit-learn} \cite{scikit}, que já inclui esses dados de maneira acessível para uso em experimentos de ML. Para visualizar um dos elementos do conjunto de dados (por exemplo, o elemento $j=0$), podemos utilizar o seguinte \textit{script} em \textit{Python}:

\noindent \begin{lstlisting}[language=Python, caption={}]
# Importando as biblioteca
from sklearn import datasets

dataset = datasets.load_iris()

j = 0 # Indice dos dados
x_j = dataset.data[j, :] # Entrada
y_j = dataset.target[j] # Alvo
print(x_j, y_j)
\end{lstlisting}
Saída:
\noindent \begin{lstlisting}[language=Python, caption={}]
[5.1, 3.5, 1.4, 0.2], [0] 
\end{lstlisting}

Após o acesso ao banco de dados, o passo subsequente consiste na inicialização dos parâmetros da rede neural. Esses parâmetros armazenam, de forma implícita, o conhecimento adquirido pelo algoritmo durante o processo de treinamento, uma vez que são utilizados para transformar as informações processadas por cada neurônio. Os ajustes realizados nesses parâmetros definem o mapeamento entre a entrada e a saída do modelo, permitindo que o Perceptron seja representado como uma função matemática da forma $g: (x_{j1}, \ldots, x_{jN}) \rightarrow y_j$.

\textbf{ Pesos e \textit{bias}:} Os principais parâmetros de uma rede neural são os pesos e {\it bias}. Os pesos ($w_i$) atribuem diferentes importâncias às entradas de cada neurônio, desempenhando um papel central na definição da relevância relativa de cada entrada para a saída do modelo. O \textit{bias} ($b$), por sua vez, é um parâmetro adicional que ajusta a saída do modelo independentemente das entradas, sendo essencial para garantir a flexibilidade e a adaptabilidade do algoritmo aos dados. No caso do Perceptron, o \textit{bias} permite deslocar a linha ou o hiperplano de decisão em espaços de maior dimensão, contribuindo para a correta classificação das entradas.

A melhor forma de inicializar os parâmetros é um campo ativo de pesquisa de ML \cite{carleo_machine_2019}, pois dependerá muitas vezes da função de ativação utilizada, a função \textit{loss}, e de características do problema tratado. Na falta de mais informação sobre o sistema, é possível gerar valores aleatórios para os pesos com distribuição normal, mas há outras opções como: inicializar todos com valores zeros, aleatoriamente com distribuição uniforme e Glorot/Xavier (para funções de ativação simétricas) \cite{glorot}. Veremos adiante que o que se chama de \textit{treinamento da rede} é uma atualização dos valores dos pesos e \textit{bias}.  

O vetor de pesos $\mathbf{w}$ e o vetor de entradas $\mathbf{x}$ podem ser escritos da seguinte forma
$$
\mathbf{w} = 
\begin{bmatrix}
   w_0,& w_1, & \cdots, & w_N
\end{bmatrix},
\quad
\mathbf{x}_j = [ x_{j0} , x_{j1},  \cdots, x_{jN} ]. 
$$

Em algumas representações, é utilizado $w_0=b$ e os dados de entrada são escritos de tal forma que $ x_{j0}=1$. Por simplicidade, adotaremos esta notação. 
Assim, podemos descrever a etapa da soma ponderada (círculo azul) ilustrada na Figura \ref{fig:perceptron}  utilizando a operação do produto escalar,
$$
\mathbf{w}^T \mathbf{x}_j = \mathbf{x}^T_j\mathbf{w}  =
\begin{bmatrix}
    x_{j0}, & x_{j1}, & \cdots, & x_{jN}
\end{bmatrix}
\begin{bmatrix}
    w_0 \\ w_1 \\ \vdots \\ w_N
\end{bmatrix} ,
$$

\noindent o super índice $T$ representa o transposto, portanto, a multiplicação matricial resultante é
$$
\mathbf{x}^T_j\mathbf{w} =
x_{j0} w_0  + x_{j1} w_1  + \cdots + x_{jN} w_N  = \sum_{i=0}^{N} w_i x_{ji}.
$$

Este produto representa uma soma ponderada das entradas de um dado $j$.  Como exemplo para ilustrar a dinâmica, vamos considerar $\mathbf{w} = [1, 0, 0.5, -0.3] $, $b = [0.5]$ e os valores de $\mathbf{x}_0$ na equação \eqref{equationX0}, que resulta em $\mathbf{w}^T \mathbf{x}_0= 6.24$. 

Este mesmo produto pode ser numericamente calculado em \textit{Python} \footnote{  Para isto usaremos da biblioteca \textit{Numerical Python (NumPy) \cite{numpycite}}, que integra uma vasta gama de funções matemáticas. Estas adições possibilitam manipulações de grandes base de dado com alta performance computacional, enquanto preserva a interface concisa e legível da linguagem \textit{Python}. Vale esclarecer aqui que podemos importar a biblioteca com uma abreviação qualquer (neste caso, por convenção de uso, ``np'').}:
\noindent 
\begin{lstlisting}[language=Python, basicstyle=\ttfamily]
import numpy as np

w = [1, 0, 0.5, -0.3]
b = [0.5]
z = np.dot(w, x_j) + b
print(z)
\end{lstlisting}
Saída:
\noindent \begin{lstlisting}[language=Python, caption={}]
[6.24]
\end{lstlisting}

\textbf{Função de ativação:}  Após a soma ponderada ser realizada, uma função não linear $f$, chamada de função de ativação, é aplicada ao resultado. As funções de ativação introduzem propriedades não lineares na rede neural, permitindo construir redes mais profundas e expressivas. Algumas comumente utilizadas incluem (veja a Tabela \ref{tab:activation_functions} em apêndice A para mais detalhes):
\begin{itemize}
    \item Sigmoide ou Logística: É uma função suave, usada principalmente na camada de saída de um modelo de classificação binária, pois sua saída varia no intervalo $[0,1]$.
    
    \item Tangente Hiperbólica (tanh): é similar à função sigmoide quanto à forma da curva, porém, varia no intervalo de $[-1,1]$, com o centro em zero. 
    
    \item Unidade Linear Retificada (ReLU): Atualmente, ReLU é uma das funções de ativação mais amplamente utilizada em redes neurais profundas devido à sua simplicidade computacional e à facilidade de cálculo de seu gradiente, sem que efeitos de não linearidade sejam perdidos. Ela reproduz a entrada se for positiva; caso contrário, produzirá zero.
    
    \item  Outros tipos: LeakyRelu, Softmax, seno, cosseno entre outras \cite{activation1,activation2}.
    
\end{itemize}

O poder de uma rede neural em modelar fenômenos complexos reside na combinação da não linearidade de sua função de ativação  com a conexão de vários neurônios (como veremos no próximo capítulo) \cite{dawid_modern_2022}. Do ponto de vista da álgebra linear, uma sequência de transformações lineares pode ser sempre reduzida a uma única transformação linear. Sem uma função de ativação não linear para quebrar essa sequência, uma rede neural com dezenas de camadas se comportaria, na prática, como uma única camada,tornando-se incapaz de aprender relações mais complexas que uma simples
regressão linear múltipla.

A introdução dessa não linearidade permite que arquiteturas profundas funcionem como aproximadores universais. Como veremos em detalhe na Seção \ref{sec:RNProfunda}, uma rede com uma composição finita de neurônios com ativações não lineares pode, em teoria, aproximar qualquer função contínua, conforme o Teorema da Aproximação Universal \ref{teo:UniversalApprox}. Embora o papel da não linearidade não seja tão evidente no Perceptron simples, cujas tarefas são, por definição, linearmente separáveis, ela é a propriedade fundamental que possibilita o aprendizado em redes mais sofisticadas. Além disso, para que o modelo possa ser treinado por métodos de otimização baseados em gradientes, a função de ativação deve ser diferenciável, isto é, ter derivada definida e finita nos pontos de interesse.

Na saída do Perceptron, obtemos a seguinte expressão:
\begin{equation}
    \hat{y}_j = f\left( \sum_{i=1}^{n} w_i x_{ji} + b  \right),
    \label{eq3}
\end{equation}
que pode ser utilizada para representar certas informações. Um dos sistemas mais simples é o classificador binário, no qual o rótulo verdadeiro $y \in \{0,1\}$ indica a classe correta, e a saída prevista $\hat{y}_j$ deve refletir uma decisão entre duas classes possíveis.
No segundo exemplo, consideramos que o rótulo verdadeiro $y$ assume valores contínuos, caracterizando um problema de regressão.

Em nosso exemplo, do banco de dados de flores, podemos classificar qual o tipo de flor a partir das características das flores. Neste caso temos três classes $\{0,1,2\}$ dadas pelas flores {\textit{setosa, versicolor, virgínica}}. Todavia, reduziremos o banco de dados para duas classes, 0 para as \textit{setosa}  e 1 para as \textit{ virgínica}, para facilitar o entendimento. Escrevendo a equação \eqref{eq3} em \textit{Python}:
\noindent \begin{lstlisting}[language=Python, caption={}]
# considerando o valor de z, obtido anteriomente
# z = [6.24]
def ativacao_sigmoide(z):
    f = 1 / (1 + np.exp(-z))
    return f

    
y_hat = ativacao_sigmoide(z)
print(y_hat)
\end{lstlisting}
Saída:
\noindent \begin{lstlisting}[language=Python, caption={}]
[0.998]
\end{lstlisting}

A escolha da função de ativação sigmoide é motivada por sua característica de limitar a saída do Perceptron ao intervalo entre 0 e 1. Esta propriedade é particularmente útil em problemas de classificação binária, pois a saída pode ser interpretada como uma probabilidade de pertencer a uma determinada classe. Por exemplo, uma saída próxima de 1 indica alta probabilidade de que a entrada pertence à classe positiva (1), enquanto uma saída próxima de 0 indica alta probabilidade de que a entrada pertence à classe negativa (0). Esta interpretação probabilística simplifica a tomada de decisões, tornando o modelo mais intuitivo e permitindo a utilização de técnicas de análise estatística. No exemplo descrito pelo código, vemos que  $\hat{y}_0 = y_{-}\text{hat} = 0.998$  indicando que para os valores de pesos inicializados, aquele dado deve pertencer à classe (1), mas se compararmos esse resultado com o valor real da saída do neurônio, notaremos que a classe desse dado deveria ser (0). Isso evidencia um erro entre o valor predito pelo Perceptron, ainda não treinado, com o valor real. Como medir esses erros e como ajustar os pesos e \textit{bias} para melhorar a previsão da saída, são tópicos das próximas subsecções.

\textbf{Função de custo (em inglês, \textit{loss function})}: também chamada de função de perda ou função de erro, é utilizada para medir o desempenho de um modelo em relação aos dados reais, quantificando o erro entre as saídas do Perceptron e os valores esperados. Essa função fornece uma métrica a ser minimizada durante o treinamento, com o objetivo de aprimorar o desempenho do algoritmo. Veremos adiante que ela também pode ser manipulada para direcionar o aprendizado do Perceptron ou de redes neurais. No nosso exemplo, utilizaremos o erro quadrático médio (em inglês, \textit{Mean Squared Error}, MSE)  que calcula a média dos erros de $m\leq M$ amostras através da fórmula:
\begin{equation}
    \mathcal{L}(y_j,\hat{y}_j) = \frac{1}{m}\sum_j^m ( y_j - \hat{y_j})^2,
    \label{eq-loss}
\end{equation}

\noindent onde $y_j$ é o rótulo verdadeiro da amostra $j$ e $\hat{y}_j$ é a previsão feita pelo modelo.  $\mathcal{L}$ representa a função custo entre  $y_j$ e  $\hat{y}_j$, sobre $m$ elementos do conjunto  total de amostras de tamanho $M$, i.e., um conjunto de $m$ linhas do {\it dataset}. No que segue iremos omitir o índice $j$ na variável $\hat{y}_j$, exceto em casos ambíguos. Escrevendo a equação \eqref{eq-loss} em \textit{Python} para $m=1$:

\noindent \begin{lstlisting}[language=Python, caption={}]
# y_hat = [0.998], y_0 = [0]
#calculo do erro quadratico com j=0
loss = (y_j - y_hat) ** 2
print(loss)
\end{lstlisting}
Saída:
\noindent \begin{lstlisting}[language=Python, caption={}]
[0.99611167]

\end{lstlisting}  

Observe que, para os valores iniciais que escolhemos, a nossa função custo está acusando um resultado maior que zero, i.e., longe do ideal. O objetivo para este dado era obter $y_j=0$, mas o valor dado pelo Perceptron foi de $\hat{y} \approx 1$. Como a função quadrática MSE tem um mínimo global em $\mathcal{L}(y_j,\hat{y}) = 0$, queremos que todos (ou a maioria) das saídas (\textit{outputs}) da rede, quando calculadas na função custo, sejam próximas deste mínimo global. 

Aqui, é importante ressaltar que nossa escolha pelo MSE serve a um propósito didático. Por ser matematicamente simples, ele nos permite ilustrar o conceito de erro de forma clara. No entanto, para problemas de classificação como este, o MSE não é a ferramenta ideal. A abordagem mais correta envolve funções de custo projetadas especificamente para classificação, sendo a Entropia Cruzada (\textit{Cross-Entropy}) uma das mais proeminentes. Esta função é mais eficaz porque penaliza previsões confiantes e erradas de forma mais acentuada, levando a um treinamento mais estável e rápido. Para não complicar este primeiro exemplo, seguimos com o MSE, mas deixamos o próximo passo como um desafio ao leitor: no material complementar, disponibilizamos um exercício que guia na implementação da Entropia Cruzada.

\textbf{Regra de aprendizado dos parâmetros:} A otimização dos parâmetros de uma rede neural é frequentemente referida como ``aprendizado'' porque esse processo de ajuste permite que a rede neural tente resolver o problema para o qual foi projetada, de maneira análoga ao processo pelo qual uma pessoa aprende a resolver um problema após estudar e praticar. Existem várias técnicas para ajustar os parâmetros de redes neurais, como algoritmos genéticos \cite{opt_genetical} e \textit{simulated annealing} \cite{simu_anneling}, que buscam soluções ideais mediante processos inspirados na evolução natural e na termodinâmica, respectivamente. No entanto, os métodos baseados em gradiente destacam-se pela sua eficácia e sucesso comprovado na otimização de algoritmos de redes neurais, sendo amplamente utilizados na prática \cite{otimization,otimization2}.

Os métodos de gradiente utilizam informações derivadas da função custo, que mede o quão bem a rede está performando em relação ao problema. A função custo pode ser visualizada, de maneira análoga, como um mapa topográfico, onde a posição inicial corresponde a um ponto elevado, como o topo de uma montanha, e o objetivo é alcançar o ponto mais baixo, que representa o valor mínimo da função. Nesse contexto, o gradiente da função custo fornece a direção de maior inclinação, permitindo que o algoritmo ``observe'' o entorno e tome decisões sobre a direção de cada passo. Ao aplicar iterativamente um método de gradiente, os parâmetros da rede neural são ajustados gradualmente para reduzir o valor da função custo, movendo-se em direção ao mínimo.

Um dos métodos de otimização de gradiente mais simples e amplamente conhecidos é o método de gradiente descendente, ou \textit{Gradient Descent} (GD) em inglês. Nesse método, calcula-se a derivada da função custo em relação a cada parâmetro que se deseja atualizar. O parâmetro é então ajustado subtraindo-se o produto da derivada com uma constante chamada taxa de aprendizado (\textit{learning rate}). Essa constante controla o tamanho do passo que o algoritmo dará na direção do gradiente. 
O ajuste dos pesos e \textit{bias}, através do método de gradiente descendente, é então dado pelas respectivas equações:
\begin{equation}
    \mathbf{w}_{\text{novos}} = \mathbf{w}_{\text{antigos}} - \eta \nabla_w \mathcal{L}
    \label{eq:SGD_w_perceptron}
\end{equation}
e
\begin{equation}
    b_{\text{novos}} = b_{\text{antigos}} - \eta \nabla_b \mathcal{L},
    \label{eq:SGD_b_perceptron}
\end{equation}

\noindent onde $\nabla_{w,b}$ representa o gradiente da função de custo em relação aos pesos $w$ ou ao \textit{bias} $b$, respectivamente, enquanto $\eta$ é a taxa de aprendizado. É importante destacar que essa taxa é definida manualmente antes do início do treinamento da rede neural e não é  geralmente ajustada durante o processo de otimização. Por ser um hiperparâmetro, $\eta$ requer uma escolha cuidadosa, pois uma taxa de aprendizado muito alta pode fazer o algoritmo oscilar ao redor do mínimo desejado ou até divergir, enquanto uma taxa muito baixa pode resultar em um treinamento extremamente lento. O valor da taxa de aprendizado também pode ser encontrado por meio de técnicas de validação cruzada ou otimização de hiperparâmetros, como busca em grade (\textit{grid search}) ou busca aleatória (\textit{random search}) \cite{Liashchynskyi}. Escolher os valores corretos para hiperparâmetros é um passo crucial para garantir o sucesso no treinamento de redes neurais, influenciando diretamente a capacidade do modelo de convergir eficientemente e evitar problemas associados ao treinamento.

Além do método básico de gradiente descendente, existem variantes desenvolvidas para melhorar a eficiência e a qualidade do treinamento de redes neurais. Entre essas variantes estão o gradiente descendente com momento (\textit{momentum}), o método de \textit{Nesterov} (\textit{Nesterov Accelerated Gradient}), o Adam (\textit{Adaptive Moment Estimation}) \cite{Gower, adamopt}. O gradiente com momento adiciona um termo de ``memória'' que ajuda a acelerar o movimento em direções que consistentemente apontam para o mínimo, enquanto o método de \textit{Nesterov} faz uma correção adicional que antecipa a direção futura. Já o Adam combina as ideias de acumulação de momentos e adapta a taxa de aprendizado para cada parâmetro, resultando em uma convergência mais rápida e estável em muitos problemas práticos.

Os gradientes da função custo, na equação \eqref{eq-loss}, em relação a $\mathbf{w}$ e $b$, aplicando a regra da cadeia de derivadas, são
\begin{align}
    \nabla_w \mathcal{L} &= -2(y_j - \hat{y}) f' \mathbf{x}_j,  \\
    \nabla_b \mathcal{L} &= -2(y_j -  \hat{y})f'.
    \label{eq-nabla}
\end{align}

A derivação da expressão \eqref{eq-nabla}, como também alguns exemplos de função custo, está em maiores detalhes no Apêndice A~\ref{sec:Apendix:CalcFuncCusto}. 
Nesse caso o termo $f'$ é determinado pela função de ativação escolhida $f$. Podemos ver na tabela 1 no apêndice as derivadas para alguns casos. Em nosso exemplo utilizamos a função sigmoide, então,
\begin{align*}
    \mathbf{w}_{\text{novos}} &= \mathbf{w}_{\text{antigos}} -2 \eta(y_j - \hat{y} ) \hat{y}(1 - \hat{y}) \mathbf{x}_j, \\
    b_{\text{novos}} &= b_{\text{antigos}} -2 \eta(y_j - \hat{y} ) \hat{y}(1 - \hat{y}).
\end{align*}

 Implementando a atualização dos pesos e \textit{bias} em \textit{Python}:
\noindent \begin{lstlisting}[language=Python, caption={}]
eta = 0.01

w = w -  2 * eta * (y_j -  y_hat) * y_hat * (1-y_hat) * x_j
b = b -  2 * eta * (y_j -  y_hat) * y_hat * (1 - y_hat)
print(w,b)
\end{lstlisting}
Saída:
\noindent \begin{lstlisting}[language=Python, caption={}]
[ 1.0002,  0.000138, 0.50005,-0.29], [0.500039]
\end{lstlisting}
Após completar o processo de aprendizagem do Perceptron com o primeiro conjunto de dados da tabela de treinamento ($j=0$), repetimos o procedimento para cada um dos dados restantes. Assim, cada linha $j$ dos dados contribuirá para ajustar os pesos. Descrevendo as etapas para os $M=150$ dados em um único \textit{script} , obteremos:  
 
\noindent \begin{lstlisting}[language=Python, caption={}]
w = [1, 0, 0.5, -0.3]
b = [0.5]
M = 150
eta = 0.01
loss = 0
for j in range(M):
    x_j = dataset.data[j, :] # Entrada
    y_j = dataset.target[j] # Alvo/rotulo
    z = np.dot(w, x_j) + b 
    y_hat = ativacao_sigmoide(z) 
    w = w - 2 * eta * (y_j - y_hat) * y_hat * (1 -y_hat) * x_j
    b = b - 2 * eta * (y_j - y_hat) * y_hat * (1 -y_hat)
    loss = loss + (y_j - y_hat) ** 2 / M
\end{lstlisting} 

Este ciclo deve ser iterado múltiplas vezes. Cada iteração em que exaurimos  todos os dados é chamada de época, ou em inglês \textit{epoch}, até que a função custo se aproxime do seu valor mínimo (neste caso, aproximadamente zero), indicando que o Perceptron aprendeu satisfatoriamente as informações fornecidas durante o treinamento. Para realizar várias iterações, adicionaremos mais um \textit{loop} usando o comando ``for''  novamente. O código completo está disponível em uma página do \textit{Github} \cite{Github}, permitindo que o leitor reproduza o experimento numérico no seu devido tempo de aprendizado.

\begin{figure}[!ht]
    \centering
    \includegraphics[width=1\linewidth]{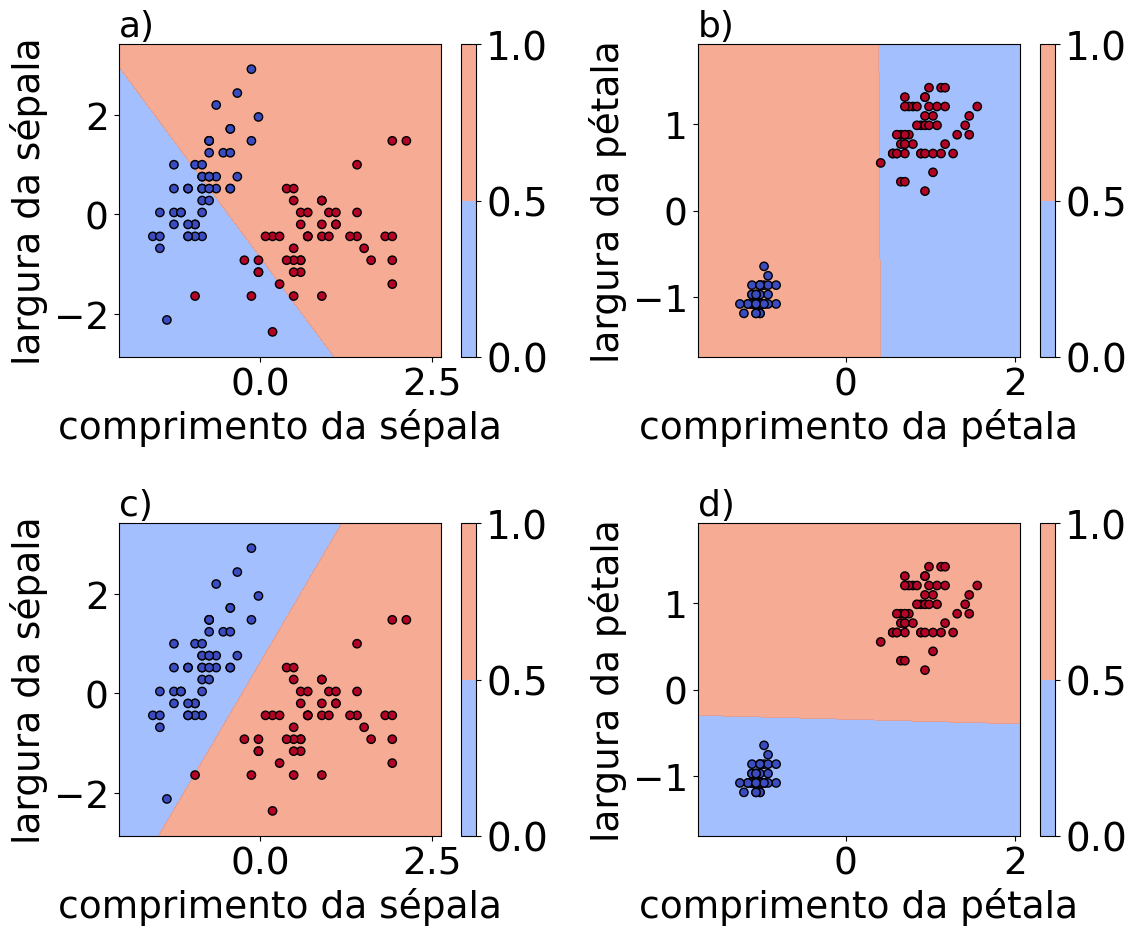}
    \caption{Evolução das Fronteiras de Decisão do Perceptron no banco de dados Íris. Os dados estão inseridos em um gráfico de dispersão, projetado em duas das quatro dimensões dos dados. A unidade de medida de ambos os eixos é dada em centímetro. Na esquerda, painéis a) e c),  escolhemos as dimensões dadas por comprimento da sépala (abcissas) e largura da sépala (ordenadas), na direita, painéis b) e d), as outras duas dimensões são o comprimento da pétala (abcissas) e largura da pétala (ordenadas). Em ambos o rótulo de aprendizado está em  codificado na cor: vermelho para classe 0 e azul para classe 1. Nos gráficos de cima, a) e b), temos a classificação dada pelo Perceptron com inicialização aleatória, já nos gráficos de baixo, c) e d), temos a classificação após a otimização. Veja que, após a otimização, o hiperplano (chamado de \textit{fronteira de decisão}, onde o Perceptron acusa o valor 0.5) separa os conjuntos de dados em ambas projeções, exceto por um ponto. }
    \label{fig:antes-depois}
\end{figure}

Na figura \ref{fig:antes-depois}, retirada do material disponibilizado no GitHub, apresentamos quatro gráficos bidimensionais. Os dois no topo com a inicialização arbitrária de pesos, e os dois na parte inferior  depois  da otimização dos parâmetros, ilustrando as fronteiras de decisão de um Perceptron aplicado ao conjunto de dados Íris. Em cada par de gráficos, o da esquerda mostra a fronteira de decisão em função do comprimento e largura da sépala, enquanto o da direita apresenta a mesma fronteira em função do comprimento e largura da pétala. Os pontos representam as amostras do conjunto de dados, onde as cores azul e vermelho indicam as diferentes classes a serem distinguidas pelo Perceptron. Para construir a fronteira de decisão do modelo de rede neural, foi configurado que todos os valores de saída da função de ativação maiores que $0.5$ são classificados como pertencentes à classe (1). Consequentemente, valores de saída menores ou iguais a $0.5$ são classificados como pertencentes à classe (0). Essa configuração define uma linha de corte que separa ambas as classes no espaço de características, permitindo que o modelo distinga entre elas com base nas saídas calculadas.

Nos gráficos \ref{fig:antes-depois}a) e \ref{fig:antes-depois}b), observamos uma considerável sobreposição das regiões coloridas,  indicando uma dificuldade significativa em separar as classes, caracterizada por uma fronteira de decisão pouco definida. Isto reflete uma baixa acurácia do modelo antes de qualquer aprendizado ocorrer. Em contraste, os gráficos  \ref{fig:antes-depois}c) e \ref{fig:antes-depois}d) mostram uma separação nítida entre as classes, com as regiões coloridas agora claramente divididas por uma fronteira de decisão bem ajustada após 100 épocas de treinamento.

\subsubsection{Exemplo Didático: Regressão}
Enquanto no exemplo anterior abordamos um problema de classificação, que envolve a categorização de dados em classes descritas por valores discretos, na regressão o objetivo é prever valores contínuos, como o preço de um produto, o peso ou a altura de uma pessoa. Em problemas de física, a regressão pode ser utilizada para prever grandezas físicas como posição, velocidade ou aceleração ao longo do tempo, sendo uma ferramenta essencial para estudar as relações entre quantidades físicas em experimentos. Neste exemplo, criaremos nossos próprios dados para ilustrar o processo de ajuste de um problema linear.  Supomos que o conjunto de dados contenha apenas duas variáveis,  uma  que represente a velocidade de um objeto sob aceleração constante e outra que represente os instantes de tempo:
\begin{align}
    \mathbf{t} &= 
    \begin{bmatrix}
        0, & 0.25, & 0.5, & 0.75, & 1
    \end{bmatrix}, \\
    \mathbf{v} &= 
    \begin{bmatrix}
       0, & 0.125, & 0.25, & 0.375, & 0.5
    \end{bmatrix}.\label{Passo0}
\end{align}

Nosso objetivo é utilizar o Perceptron como uma função que encontre a relação dos dados de entrada com as saídas, para depois prever a velocidade (saída) conhecendo o tempo (entrada).  Observe que os problemas de regressão também são considerados aprendizado supervisionado, pois os rótulos de saída $y_i$ neste caso, a velocidade, são fornecidos durante o treinamento. Com $\mathbf{t}$ sendo os dados de entrada e $\mathbf{v}$ os pontos experimentais que desejamos ajustar. Nesse problema, cada item de $\mathbf{v}$ é equivalente a $y_i$  e $\mathbf{t}$ representa $x_{ji}$ descrito na equação \eqref{eq-1} com $N=1$ e $M=5$. Assim, tendo conhecimento dos dados de entrada e saída, o próximo passo é gerar o peso e \textit{bias} aleatórios, no qual utilizaremos:
\begin{align}
    \mathbf{w} = [0.497], \mathbf{b} = [-0.138].\label{Passo1}
\end{align}

A seguir, realizamos o processo de propagação em inglês \textit{forward}, que consiste na propagação dos dados de entrada utilizando a equação \eqref{eq3} para cada valor de $t$.  Antes de prosseguir, é importante esclarecer um conceito importante em redes neurais: o \textit{batching}. Essa técnica consiste em dividir o conjunto de dados em pequenos subconjuntos chamados de \textit{batches}, ou lotes, em português. Durante o processo de treinamento de uma rede neural, em vez de calcular a função custo para cada exemplo individualmente, o \textit{batching} permite que o cálculo seja realizado para um grupo de exemplos de uma só vez. Em seguida, a otimização dos parâmetros do modelo é realizada com base na média ou soma dos gradientes computados para todos os exemplos do \textit{batch}. Esta abordagem não apenas melhora a eficiência computacional do treinamento, mas também proporciona uma melhor generalização, ao reduzir o ruído estocástico associado a cada atualização de parâmetro. Assim, o \textit{batching} é amplamente utilizado para acelerar a convergência do modelo e estabilizar o processo de otimização, especialmente em problemas com grandes conjuntos de dados. Descreveremos em detalhes alguns passos matemáticos que acontecem no Perceptron, visando auxiliar o leitor a entender todo o processo. 

Iniciamos com o cálculo do produto interno, sobre o \textit{batch} de 5 pontos em $t$:
\begin{align*}
    &\mathbf{w}\mathbf{t} + b = [-0.138  , -0.01375,  0.1105 ,  0.23475,  0.359]. 
\end{align*}

Em seguida, aplicamos a parte não linear com a função ReLU (ver apêndice A) sobre todo o \textit{batch} :
\begin{align*}
    f(\mathbf{w}\mathbf{t} + b) 
    & = [0, 0, 0.1105 ,  0.23475,  0.359].
\end{align*}

O resultado de $f(\mathbf{w}\mathbf{t} + b)$ são os  valores previstos para velocidade $\hat{y}$, que utilizaremos para calcular o erro em relação a $\mathbf{v}$ com a função custo \eqref{eq-loss}, que resulta em
\begin{align*}
    \mathcal{L} & = \sum_{\text{batch}} (\mathbf{v} - \hat{y})^2=0.0746.
\end{align*}

Para ajustar os novos pesos e \textit{bias}, utilizamos a equação \eqref{eq-nabla}. A derivada da função ReLU segue a regra: $f'(x) = 1$ se $x > 0$, e $f'(x) = 0$ se $x \leq 0$, de modo que
\begin{align*}
      f'(\mathbf{w}\mathbf{t} + b) &= f'([0, 0, 0.1105 ,  0.23475,  0.359])= [0, 0, 1, 1, 1].
\end{align*}

Assumindo que $\eta = 0.05$, temos:
\begin{align}
    2\eta(\mathbf{v}- \hat{y})f'_{j}\mathbf{t} &= [0, 0, 0.0069, 0.0105, 0.0141]\label{Passo3}.
\end{align}

Tomando a média do \textit{batch}, obtemos
\begin{equation*}
     \eta \sum_{\text{batch}} \nabla_w \mathcal{L} = 0.0063.
\end{equation*}

Utilizando os passos \eqref{Passo0}, \eqref{Passo1}, \eqref{Passo3} na equação \eqref{eq-nabla}, podemos obter os novos pesos e \textit{bias}:
\begin{align*}
    \mathbf{w}_{\text{novos}}  &= \mathbf{w}_{\text{antigos}} - 0.0063= 0.4906\\
    b_{\text{novos}}  &= b_{\text{antigos}} - 0.008= -0.1464.
\end{align*}

\begin{figure}[h]
    \centering
    \includegraphics[width=1\linewidth]{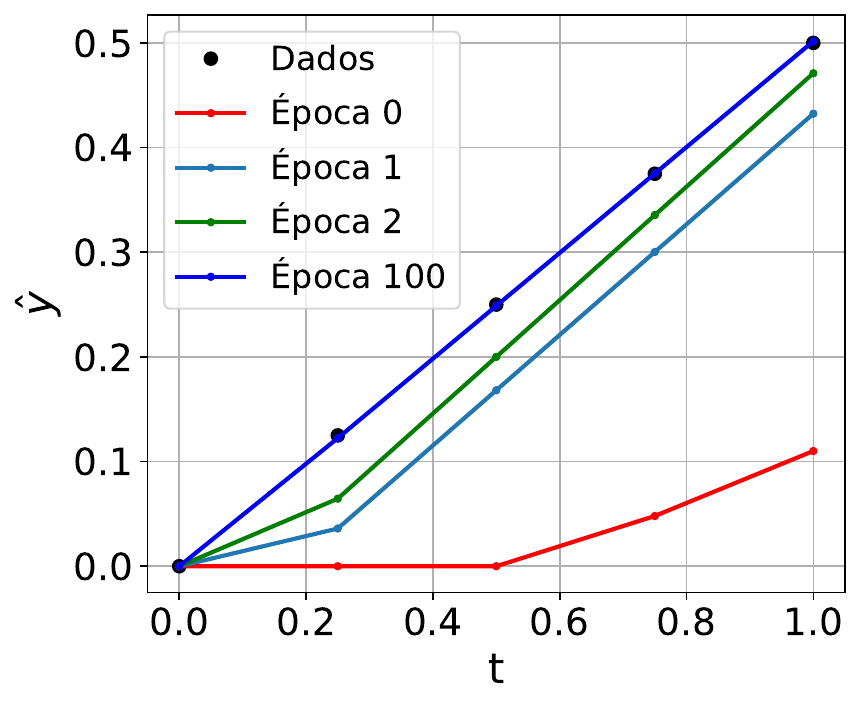}
    \caption{Comportamento do Perceptron para o problema de regressão ao longo de diferentes épocas de treinamento. No eixo horizontal temos os instantes de tempo ($t$) e no eixo vertical a predição da velocidade ($\hat{y}$) feita pelo Perceptron. Os pontos pretos representam os dados de treino, enquanto as linhas vermelha, verde e azul correspondem às estimativas do modelo nas épocas 0, 1,2 e 100, respectivamente.} 
    \label{fig:regressao}
\end{figure}

Com isso, completamos a primeira época e esperamos obter uma função custo com valor menor que o anterior. Podemos verificar isso repetindo o processo anterior mais uma vez.
\begin{align*}
    \hat{y}&=f(\mathbf{w}\mathbf{t} + b) = [0, 0.036, 0.178, 0.32, 0.46],\\
    \mathcal{L} &=\sum_{\text{batch}} (\mathbf{v} - \hat{y})^2 = 0.0643.
\end{align*} 
Comparando o valor da função custo dos novos pesos com o anterior, notamos que houve uma redução de $0.0103$, isso indica que o Perceptron está começando a aprender as informações dos dados. O processo pode ser repetido várias vezes para reduzir progressivamente a função custo. Em nosso material complementar no \textit{Github}, apresentamos a continuação deste exemplo, onde executamos o algoritmo por mais 98 épocas, totalizando 100 épocas. Os resultados deste teste são apresentados na Figura \ref{fig:regressao}, onde alcançamos  o valor de erro de $1.059\times 10^{-7}$.

Na Figura \ref{fig:regressao} temos o resultado da convergência do modelo de regressão ao longo das épocas de treinamento, com a linha do modelo se ajustando progressivamente aos dados observados. Este comportamento evidencia a capacidade do Perceptron em aprender e ajustar-se a um conjunto de dados simples por meio de sucessivas atualizações de seus parâmetros.
\subsection{Redes neurais profundas}
\label{sec:RNProfunda}
Uma rede neural artificial é composta por um conjunto de neurônios artificiais interconectados, cuja organização define a estrutura do modelo. A maneira como esses neurônios são interligados é conhecida como arquitetura da rede. Existem diversas formas de organizar essas conexões. Uma configuração comum envolve agrupar neurônios em unidades funcionais conhecidas como camadas. Cada camada é composta por neurônios que operam de forma independente, mas que processam informações antes de transmitir suas saídas para a camada seguinte. 

Uma das arquiteturas mais conhecidas no ML é o \textit{Perceptron Multicamadas} (MLP, do inglês \textit{Multi-Layer Perceptron}), um tipo de rede neural \textit{feed-forward} densa. Redes \textit{feed-forward} distinguem-se pela ausência de conexões recorrentes, ou seja, as saídas de uma camada são utilizadas exclusivamente como entradas para a camada seguinte. Esse fluxo unidirecional de informações ocorre da camada de entrada até a camada de saída. Especificamente, os neurônios de uma camada $n$ fornecem entradas para os neurônios da camada subsequente, $n+1$, sem retornos ou ciclos no processo. Essa arquitetura é amplamente utilizada devido à sua simplicidade e eficácia em tarefas de classificação e regressão, onde uma relação direta entre entradas e saídas é desejada.

\begin{figure}[htbp]
    \centering
    \includegraphics[width=1.0\linewidth]{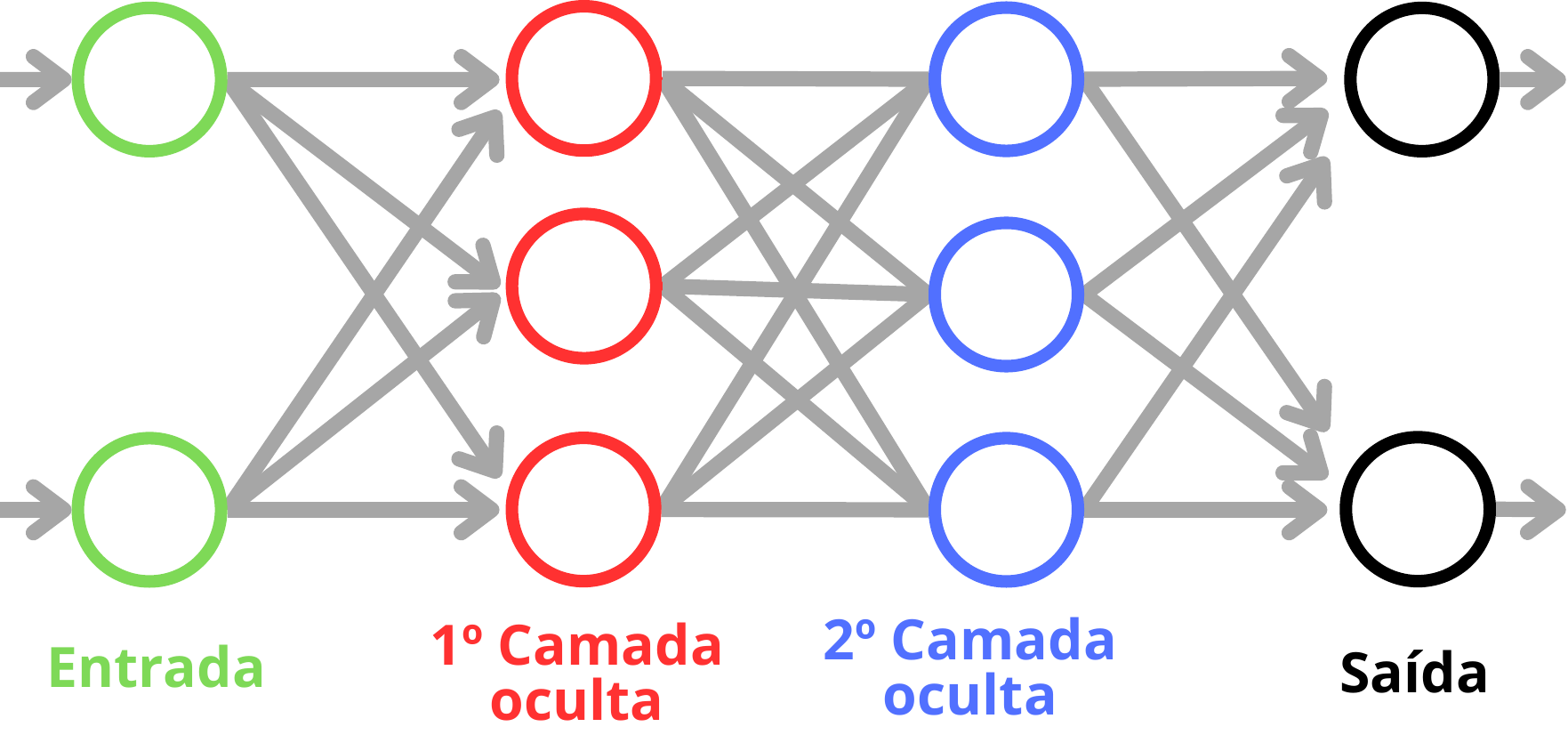}
    \caption{ Representação gráfica de uma configuração específica de rede neural utilizando grafo, composta por 2 entradas, 2 camadas ocultas com 3 neurônios cada e 2 saídas. Os círculos verdes, representam as entradas $x_{ji}$ e os pretos, as saídas $\hat{y}_{i}$. As camadas do meio, compostas pelos neurônios em vermelho e azul, são as camadas ocultas.}
    \label{fig:NeuralNetwork}
\end{figure}

A representação gráfica da arquitetura MLP está ilustrada na Figura \ref{fig:NeuralNetwork}, lida da esquerda para a direita: as entradas são representadas pelas linhas originadas da ponta esquerda da imagem, conectadas aos nodos em verde, $x_{ji}$ na notação do capítulo anterior. As entradas estão conectadas com a primeira camada (em vermelho), onde cada nodo é um Perceptron, cuja saída será usada de argumento na segunda camada de neurônios (em azul), os quais, por sua vez, se conectam à camada final de neurônios da direita (também em verde). Estes, finalmente, correspondem à saída da rede. Cada nodo pode ter a sua função de ativação específica, assim como as conexões neurônio a neurônio podem ser arbitrariamente determinadas, desde que seja sempre com as camadas vizinhas, somente. As camadas intermediárias entre as de entrada e saída (neste exemplo: vermelha e azul) são chamadas de \textit{camadas ocultas}.

Por simplicidade, utilizaremos o termo ``redes neurais'' ao nos referirmos especificamente às redes neurais artificiais com arquitetura \textit{feed-forward}. No entanto, é importante que o leitor compreenda que o campo das redes neurais é vasto e inclui uma grande diversidade de arquiteturas, cada uma projetada com características específicas para atender a diferentes tipos de problemas. Entre essas outras arquiteturas, destacam-se as redes neurais recorrentes (RNNs), que são particularmente eficazes para lidar com dados sequenciais, como séries temporais  \cite{NathanKutz2017,Rumelhart,LeCun} e processamento de linguagem natural, muitas vezes por meio do uso de LSTM (\textit{Long Short-Term Memory}) \cite{Hochreiter}; as redes neurais convolucionais (CNNs), que exploraremos mais aprofundadamente nas seções que seguem, são amplamente utilizadas em tarefas de visão computacional, como reconhecimento de imagens e detecção de objetos, devido à sua capacidade de extrair características hierárquicas de dados espaciais \cite{Marquardt2021}; e os \textit{Transformers}, que revolucionaram o campo do processamento de linguagem natural e também têm sido aplicados com sucesso em outras áreas devido à sua habilidade de capturar dependências de longo alcance em dados sequenciais sem a necessidade de processamento recorrente \cite{transforme}. Há, além dessas,  outras arquiteturas como \textit{Autoencoders}, \textit{Deep Belief Networks}, Redes Adversárias Generativas (GANs), Máquinas de Boltzmann aliadas a métodos como Aprendizado Profundo Bayesiano, que têm contribuído significativamente para o avanço da área \cite{Michelucci,Alom}. Uma representação pictográfica das muitas possíveis arquiteturas pode ser encontrada em \cite{NathanKutz2017}. Recomendamos a leitura de Herberg et al. \cite{Herberg} para aqueles interessados em explorar essas arquiteturas detalhadamente. 

Recentemente, John J. Hopfield e Geoffrey Hinton foram laureados com  o Prêmio Nobel de Física de 2024, concedido  por suas contribuições fundamentais ao ML com redes neurais artificiais. Hopfield desenvolveu redes capazes de armazenar e reconstruir padrões de dados usando princípios da física conhecida como Redes de Hopfield \cite{hopfield}, enquanto Hinton desenvolveu a máquina de Boltzmann, que reconhece padrões em dados utilizando ferramentas da física estatística. Essas inovações foram a base para avanços significativos em inteligência artificial e aprendizado profundo.

A escolha da arquitetura de uma rede neural deve ser cuidadosamente considerada e está intimamente ligada ao problema específico que se deseja resolver, às características dos dados envolvidos, e aos requisitos de desempenho e eficiência do modelo. 
De maneira geral, o objetivo de uma rede neural é aproximar uma função ideal $g: U \rightarrow V$,  não necessariamente conhecida, onde $U,V$ são espaços vetoriais. Assim, sendo $\mathbf{x} \in U$ um elemento do conjunto de dados e $\boldsymbol{W},\boldsymbol{b} \in \mathbb{R}^d$ vetores dos parâmetros, então queremos que $\textit{NN}: U~ \times ~ \mathbb{R}^d \rightarrow V$ satisfaça
\begin{equation}
     \textit{NN}(\mathbf{x}; \boldsymbol{W},\boldsymbol{b}) \approx g(\mathbf{x}).
\end{equation}

Por exemplo, em regressão, podemos ter uma entrada $\mathbf{x}$ associada a um valor $g(\mathbf{x})$, onde a rede neural encontrará a função  $g(\mathbf{x})$  através dos valores dos parâmetros $\boldsymbol{W}, \boldsymbol{b}$ resultando na melhor aproximação dessa função, $\textit{NN}(\mathbf{x}; \boldsymbol{W},\boldsymbol{b})$.

Vamos detalhar matematicamente o que descrevemos sobre a arquitetura \textit{feed-forward} considerando os neurônios entre camadas totalmente conectados, pictorialmente representados na Figura \ref{fig:NeuralNetwork}. A rede transforma a entrada \(\mathbf{x}\) em saída  \( \textit{NN}(\mathbf{x}; \boldsymbol{W},\boldsymbol{b})\) usando a seguinte estrutura (considerando a primeira camada como a vermelha na figura, e a entrada como os nós verdes).

\begin{align*}
    \textbf{ Entrada}&= \mathbf{x}, \\
    \textbf{1ª Camada}&= \Big(f^{[1]}_1 ( \sum_{i=1}^{N_0}{w^{[1]}_{i1} x_i} + b^{[1]}_1), \ldots, \\  &\qquad \ldots, f_{N_1}^{[1]}(\sum_{i=1}^{N_0} w^{[1]}_{iN_1} x_i + b^{[1]}_{N_1}) \Big),\\
    \textbf{ 2° Camada }&= \Big( f^{[2]}_1 ( \sum_{i'=1}^{N_1} w^{[2]}_{i'1} f^{[1]}_{i'} (\ldots) + b^{[2]}_1   ), \ldots \\
    &\qquad \ldots,   f^{[2]}_{N_2} ( \sum_{i'=1}^{N_1} w^{[2]}_{i'N_2} f^{[1]}_{i'} (\ldots) + b^{[2]}_{N_2}   ) \Big),\\ 
     &~\vdots  \\
\end{align*}
\begin{align*}
    \textbf{ Saída }&= \Big( f^{[n]}_{1} (\sum_{i'=1}^{N_{(n-1)}} w^{[n]}_{i'1} f^{[n-1]}_{i'} (\cdots) + b^{[n]}_{1}   ), \ldots, \\
    & \qquad \ldots, f^{[n]}_{N_n} (\sum_{i'=1}^{N_{(n-1)}} w^{[n]}_{i'{N_n}} f^{[n-1]}_{i'} (\cdots) + b^{[n]}_{N_n} ) \Big), \\
    \textbf{ Saída }&= \textit{NN}(\mathbf{x}; \boldsymbol{\theta}),  
\end{align*}
onde $f^{[n]}_{k_n}$ e \(b^{[n]}_{k_n}\) são a função de ativação e vetor de \textit{bias} da $n$-ésima camada associado ao $k$-ésimo neurônio desta camada $n$, respectivamente, e \(w^{[n]}_{lk_{n}}\) é o peso que liga o $l$-ésimo neurônio da camada $n-1$ ao $k$-ésimo neurônio da camada $n$.

O leitor pode estar se perguntando qual é a garantia de que esta longa composição iterada de somas pesadas de composições aproxima realmente a função desejada $g(\mathbf{x})$. A resposta a essa pergunta é afirmativa, dado o seguinte teorema \cite{dawid_modern_2022, KAN2024}:

\begin{teorema}[Aproximação Universal: {Kolmogorov-Arnold}]
    Seja $g: \mathbb{R}^N \rightarrow \mathbb{R}$ uma função contínua, e sejam $\varphi_q,\phi_{q,p}:\mathbb{R} \rightarrow\mathbb{R}$, funções não lineares, com $q\in \{0,\ldots,2N\}$ e $p \in \{0,\ldots,N\}$. Então $g$ é aproximada por 

    \begin{equation}
        f(\mathbf{x}) = \sum_{q=0}^{2N} \varphi_q\Big(\sum_{p=0}^N \phi_{q,p}(x_p)\Big)  \approx g(\mathbf{x}),
    \end{equation}
    onde $\mathbf{x} \in \mathbb{R}^N$.
    \label{teo:UniversalApprox}
\end{teorema}

Isto é, podemos representar qualquer função contínua com um número polinomial $\mathcal{O}(N^2)$ de funções não lineares de uma variável. O leitor deve ter percebido como a função do teorema é similar a uma rede neural com duas camadas e $b^{[n]}=0, \forall n$ . De fato, com somente duas camadas, qualquer função contínua pode ser aproximada, caso tenha neurônios suficientes. Pragmaticamente, é boa prática de ML trabalhar com um número maior de camadas menores, melhorando o desempenho computacional \cite{Marquardt2021}. 

Visto o teorema anterior, poder-se-ia alegar que ML e Redes Neurais são \textit{somente} métodos de aproximação de curvas e dados. Esta afirmação não está de todo errada, porém é uma generalização apressada. Pense na seguinte analogia: ``Física de muitos corpos quânticos é somente a equação de \textit{Schrödinger} em espaços de alta dimensão'', sabemos que, na prática, há muitos novos fenômenos nesta área, que requerem novas técnicas que estão ausentes em mecânica quântica de poucas partículas, o que justifica que a afirmação acima é muito redutiva. O mesmo pode ser dito de redes neurais \cite{Marquardt2021}.

Para aplicações práticas, construir redes neurais profundas somente com bibliotecas como \textit{NumPy}, como fizemos para o Perceptron na Seção \ref{sec:perceptron}, é uma tarefa custosa. Devida à complexidade e tamanho das redes neurais, é de interesse que usemos bibliotecas prontas em Python, como \textit{PyTorch} \cite{pytoch}, \textit{TensorFlow} \cite{tensorflow} e \textit{Jax} \cite{jax}, com o intuito de facilitar o uso, já que muitas das operações típicas de treinos em redes neurais (as quais trabalhamos na parte anterior do texto), já vêm em funções pré-definidas nestas bibliotecas. Além disso, um recurso fundamental dessas bibliotecas é a diferenciação automática, essencial no treinamento de redes neurais. Trata-se de uma técnica eficiente para calcular gradientes com exatidão e de forma automatizada, eliminando a necessidade de derivação manual ou de aproximações numéricas, como as diferenças finitas \cite{autodiff}. A diferenciação automática automatiza esses cálculos usando a regra da cadeia usual para derivadas, facilitando a aplicação de algoritmos como o gradiente descendente. Isso acelera o desenvolvimento, garante maior precisão, e preserva estruturas diferenciáveis,  sendo uma das razões pelas quais essas ferramentas são amplamente adotadas no ML. 

O leitor pode encontrar no link do nosso \textit{Github} em \cite{Github} o material complementar a este artigo, onde há \textit{jupyter notebooks} com os devidos passos a passos, explicando como criar uma rede neural usando a biblioteca \textit{PyTorch}. Para introduzir a biblioteca, preparamos um \textit{jupyter} específico aplicado ao mesmo problema de classificação de flores (\textit{Iris dataset}) utilizado anteriormente.

\begin{figure}[!ht]
    \centering
    \includegraphics[width=1\linewidth]{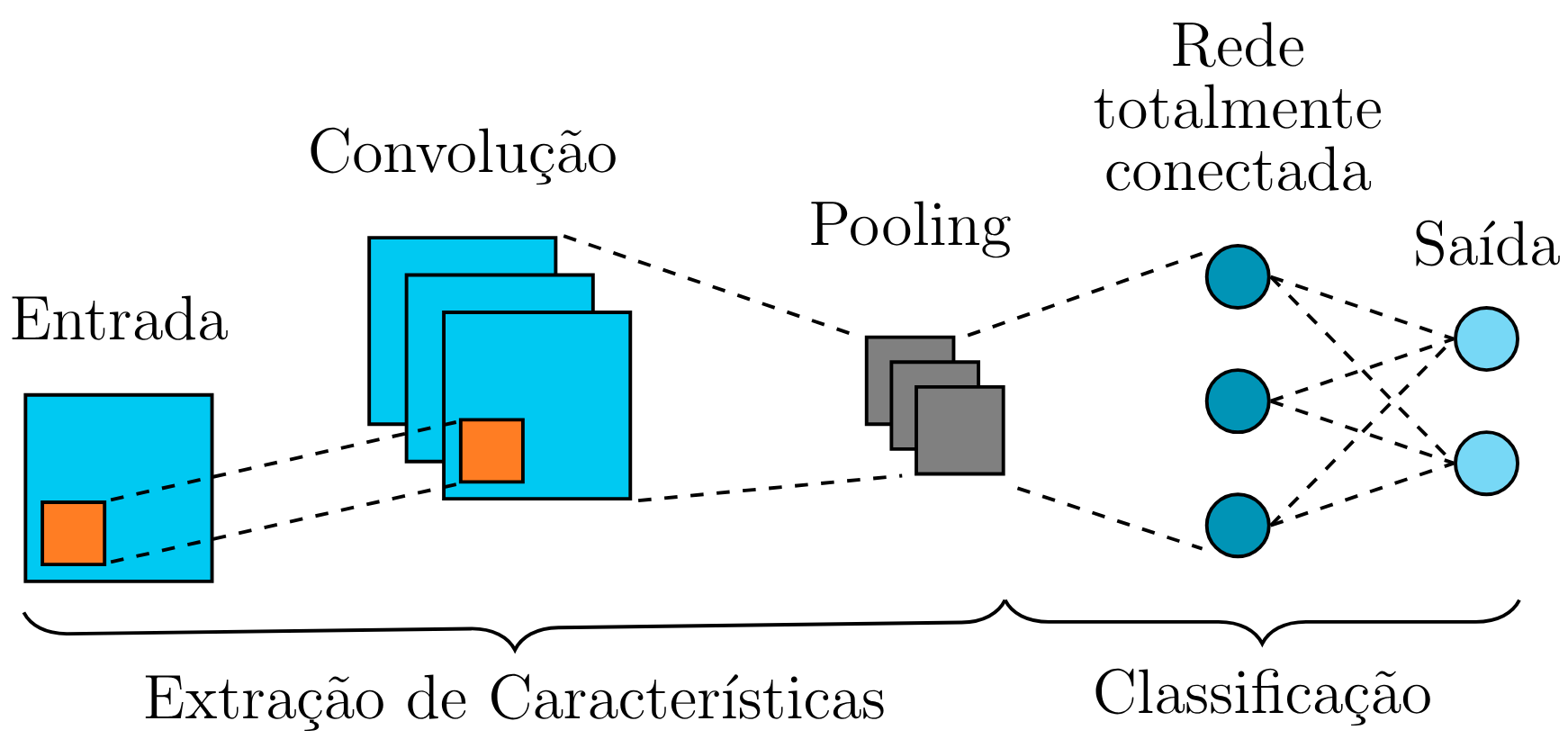}
    \caption{Representação pictórica de uma Rede Neural Convolucional. Dada uma imagem (uma matriz), submatrizes (em laranja)
    deste dado servem de argumento para a camada de filtros. Cada filtro serve como uma convolução desta submatriz em que cada kernel desta convolução é específico para abstrair alguma característica da imagem. O resultado desta camada é processado por uma camada de \textit{pooling} que reduz a dimensão das matrizes resultantes. Finalmente, as matrizes reduzidas são argumentos para uma rede totalmente conectada que aprende padrões baseados nestes dados de dimensão menor. }
    \label{fig:cnn}
\end{figure}

\textbf{Redes Neurais Convolucionais} 

As redes neurais convolucionais, do inglês \textit{Convolutional Neural Networks} (CNNs), representadas na Figura \ref{fig:cnn}, se diferenciam das redes totalmente conectadas por dois conceitos fundamentais: a localidade das conexões e o compartilhamento de parâmetros. A localidade refere-se ao fato de que um neurônio, ou um conjunto de neurônios, está conectado apenas a um subgrupo de neurônios da camada anterior, em vez de estar conectado a todos os neurônios dessa camada, como ocorre nas redes totalmente conectadas. Essa característica permite que as CNNs capturem padrões locais, como bordas e texturas em imagens, mantendo uma representação mais compacta dos dados.

O compartilhamento de parâmetros, por outro lado, significa que todos os neurônios de convolução dentro de uma mesma camada, denominados filtros,  utilizam o mesmo conjunto de parâmetros, que inclui pesos e \textit{bias}. Esses neurônios se distinguem apenas pela região da entrada à qual estão conectados, possibilitando a detecção eficiente de características similares em diferentes partes da entrada, como uma imagem. Devido a essa propriedade, as CNNs são altamente eficientes para tarefas onde a posição relativa de uma característica é mais importante do que sua localização exata, como em reconhecimento de padrões visuais e visão computacional \cite{bojarski2017,NIPS2012_c399862d}.

Além disso, é comum em arquiteturas convolucionais aplicar múltiplos filtros em paralelo sobre os mesmos valores de entrada. Cada filtro possui seu próprio conjunto de parâmetros que é compartilhado entre os neurônios que o compõem. Esta abordagem permite que a rede extraia diferentes tipos de características, como bordas horizontais, verticais ou texturas complexas, simultaneamente. 

Quando usamos a palavra `convolução' na Física, geralmente estamos expressando a ideia de que, se uma partícula em $x$ é funcionalmente representada por $f(x)$, e a maneira como ela interage com seu ambiente (ponto a ponto) é regida por uma função (ou um campo) $G(x)$, então, pode-se representar a atuação das redondezas (digamos nos pontos $x'$ vizinhos) sobre a partícula, pela convolução $A(x) = \int G(x-x')f(x')dx'$ \cite{Marquardt2021}. Em geral, a função $G(x-x')$, chamada de \textit{núcleo} ou \textit{kernel}, preserva uma série de propriedades que sabemos ser importantes na física, como, por exemplo, causalidade. A mesma ideia é aplicada em CNNs: O núcleo da convolução, o \textit{filtro},  fornece informação de como um píxel em uma imagem está relacionado com seus vizinhos. Neste caso, o uso do termo ``filtro'' foi herdado da área de pesquisa em visão computacional \cite{Lindeberg1998}.

\begin{figure} [ht]
    \centering
    \includegraphics[scale=0.25]{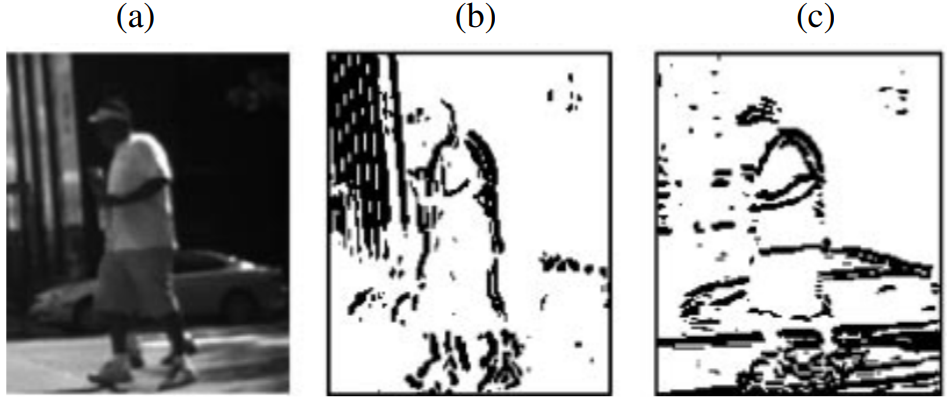}
    \caption{
    Exemplos de usos de convoluções para detecção de bordas. (a) Imagem original, (b) sob o filtro $G_x$, (c) sob o filtro $G_y$. Filtros estão explicitamente expressos na equação (\ref{eq:ExempKernel}). Fonte  \cite{EdgeDetec2008}. }
    \label{fig:FiltrosCNN}
\end{figure}

Veja na Equação (\ref{eq:ExempKernel}) dois exemplos \cite{EdgeDetec2008} de filtros usados para encontrar bordas nas imagens, nas direções $x$ e $y$, e seus efeitos em imagens na Figura \ref{fig:FiltrosCNN} (estes filtros são chamados em computação visual de {\it filtros de detecção de bordas} ou filtros {\it de Sobel} \cite{LindebergScaleSpace1993}) .

\begin{equation}
    G_x = \frac{1}{3}
    \begin{bmatrix}
    -1 & 0 & 1 \\
    -1 & 0 & 1 \\
    -1 & 0 & 1 \\
    \end{bmatrix}, \qquad 
    G_y = \frac{1}{3}
    \begin{bmatrix}
    -1 & -1 & -1 \\
    0 & 0 & 0 \\
    1 & 1 & 1 \\
    \end{bmatrix}.
    \label{eq:ExempKernel}
\end{equation}

Diferentemente da visão computacional clássica, onde filtros de imagem são projetados manualmente para tarefas específicas, as redes convolucionais aprendem os filtros ideais de forma autônoma. No início do treinamento, os parâmetros que definem os filtros são inicializados aleatoriamente. À medida que o modelo é otimizado para minimizar a função de custo, esses pesos são ajustados iterativamente. Como resultado, a própria rede descobre o conjunto de filtros mais eficaz para extrair informações relevantes. Tipicamente, as camadas iniciais convergem para filtros que detectam características fundamentais, como bordas, texturas e gradientes de cor. Camadas subsequentes aprendem a combinar essas características simples para identificar padrões mais complexos, como partes de objetos, tornando o processo de extração de características mais poderoso e adaptável do que um conjunto de filtros pré-definidos.

É prática comum na literatura que as camadas convolucionais sejam seguidas de camadas de {\it pooling} e posteriormente camadas densas. Esta arquitetura mista está representada na Figura \ref{fig:cnn}. As camadas de \textit{pooling} desempenham um papel na redução das dimensões espaciais da entrada, ao mesmo tempo que preservam as características mais importantes da imagem ou do dado de entrada. Essa redução  da dimensionalidade torna a rede neural mais eficiente em termos de custo computacional, viabilizando o uso de redes mais profundas, e torna a performance da rede   menos sensível a pequenas variações na posição das características detectadas. Já as camadas densas servem para o uso final específico do que se queira deduzir das imagens analisadas, seja classificação, remoção de ruídos, etc \cite{NathanKutz2017}. Isso se deve ao fato de que as camadas densas reúnem e avaliam correlações entre os padrões extraídos pelas camadas convolucionais, permitindo inferências sobre características globais da imagem processadas. Há uma inspiração biológica para esta escolha, que está fora do escopo deste trabalho, mas que pode ser encontrada em \cite{FukushimaCNN1980}.

O \textit{pooling} é uma operação que considera um conjunto de elementos, como uma matriz de tamanho $K\times K$, e aplica uma regra de redução que gera uma matriz de tamanho menor, por exemplo, $\frac{K}{2} \times \frac{K}{2}$ \cite{NathanKutz2017}. Há várias formas de realizar essa redução, cada uma com suas próprias características e vantagens. Vamos considerar dois exemplos:

1) Dado um bloco de pixels de uma imagem maior, $B_{4\times 4}$, i.e., uma submatriz da matriz de um dado de entrada, então
\begin{align*}
    \textit{pooling} (B_{4\times 4}) = M_{2 \times 2},  
\end{align*}
onde
\begin{align*}
     M_{ij} = \frac{1}{4} \sum_{k=i,l=j}^{i+2,j+2} B_{kl}.
\end{align*}

\noindent i.e., cada elemento $M_{ij}$ é a média dos elementos de uma submatriz $2\times 2$ de $B_{4\times 4}$. 

2) Podemos fazer o mesmo, mas agora tomando o maior dos elementos $M_{ij} = \max_{k=i,l=j}^{i+2,j+2}(B_{kl}) $. Veja que, após tal operação, o tamanho do dado de entrada está agora comprimido em uma matriz menor. Reduzir a dimensionalidade dos dados diminui o  tempo de processamento da rede, e logo acelera o processo de aprendizado também, porém sempre há perda de informação nesses processos, logo um balanço entre as duas operações deve ser encontrado.

\subsection{Treinamento em Redes profundas}

No contexto de redes neurais profundas, o algoritmo de treinamento preserva, em essência, a mesma forma matemática apresentada nas equações \eqref{eq:SGD_w_perceptron} e \eqref{eq:SGD_b_perceptron}; contudo, os pesos e \textit{bias} passam a ser denotados por $w^{[k]}_{il}$ e $b^{[k]}_l$, respectivamente.
Veja que agora os pesos e \textit{bias} necessitam de um índice a mais que no caso do Perceptron, o qual indica a camada de que o peso advém. Queremos otimizar os pesos de qualquer camada arbitrária, mas sabemos somente o resultado da função custo, a qual é calculada pela saída da rede, e esta, por sua vez, é uma composição de todas as operações internas na rede, em uma ordem bem definida. Felizmente, há uma ferramenta muito bem conhecida para saber a derivada de funções compostas, que é a regra da cadeia. Lembre-se que uma saída do $l'$-ésimo neurônio $y^{[n]}_{l'}$ da rede de $n$ camadas tem derivada com relação à $w^{[k]}_{il}$ dada por $\frac{d}{dw^{[k]}_{il}} y^{[n]}_{l'} $. Escrevendo o resultado da soma com os respectivos pesos: $z^{[n]}_{l'}= \sum_{i'=1}^{N_{(n-1)}} w^{[n]}_{i'l'}  y^{[n-1]}_{i'}+ b^{[n]}_{l'} $, i.e., calculada no neurônio $l'$ da camada $n$, temos, pela regra da cadeia
\begin{align*}
\frac{d}{dw^{[k]}_{il}} y^{[n]}_{l'} &= \frac{d}{dw^{[k]}_{il}} \Big( f^{[n]}_{l'} (z^{[n]}_{l'})\Big)   \\
&=\frac{d}{dz^{[n]}_{l'}} f^{[n]}_{l'} (z^{[n]}_{l'})   \cdot \sum_{i'=1}^{N_{(n-1)}}  w^{[n]}_{i'l'} \frac{d}{dw^{[k]}_{il}} \Big( y^{[n-1]}_{i'} \Big). 
\end{align*}

Logo, a derivada $\frac{d}{dw^{[k]}_{il}} y^{[n]}_{l'} $ depende da derivada de $\frac{d}{dw^{[k]}_{il}} y^{[n-1]}_{i'} $, o que mostra que temos uma estrutura recursiva neste cálculo. Ademais, como os índices $i'$ se repetem na soma $$\frac{d}{dw^{[k]}_{il}}z^{[n-1]}_{l'} = \sum_{i'=1}^{N_{(n-1)}}  w^{[n]}_{i'l'} \frac{d}{dz^{[n-1]}_{i'}} f^{[n-1]}_{i'} (z^{[n-1]}_{i'}) \frac{d}{dw^{[k]}_{il}} \Big( z^{[n-1]}_{i'}\Big),$$ podemos definir uma matriz 
\begin{equation}
    M^{[n,n-1]}_{i',l'} = w^{[n]}_{i'l'} \frac{d}{dz^{[n-1]}_{i'}} f^{[n-1]}_{i'} (z^{[n-1]}_{i'}).
\end{equation}

Então, podemos escrever, recursivamente
\begin{equation}
    \frac{d}{dw^{[k]}_{il}}z^{[n-1]}_{l'} =  M^{[n,n-1]}M^{[n-1,n-2]}\ldots M^{[k+1,k]}\frac{d}{dw^{[k]}_{il}}z^{[k]}_{l}.
    \label{eq:matrizcompositionbackprop}
\end{equation}

Veja que este produto é uma multiplicação de matrizes usual, logo o treino da rede neural é feito por um algoritmo linear, apesar da rede, em si, ser não linear. Então, para o último termo
\begin{align*}
    \frac{d}{dw^{[k]}_{il}}z^{[k]}_{l}   &= \sum_{i'=1}^{N_{k}} \frac{d}{dw^{[k]}_{il}} \Big( w^{[k]}_{i'l'}  y^{[k-1]}_{i'}\Big) \\
    &= \sum_{i'=1}^{N_{k}} \delta_{i'i}\delta_{l'l} y^{[k-1]}_{i'} \\
    &= y^{[k-1]}_{i}.
\end{align*}

Similarmente,
\begin{align*}
        \frac{d}{db^{[k]}_l}z^{[k]}_{l}   &= 1.
\end{align*}

Este processo de propagação do erro para trás é conhecido como \textit{backpropagation}. Sua notável eficiência computacional é o que torna o treinamento de redes profundas prático. Em vez de calcular o gradiente da função de custo em relação a cada peso de forma independente, um processo que seria proibitivamente lento, o algoritmo emprega uma técnica mais inteligente. Durante a propagação direta (\textit{forward pass}), os valores de ativação de cada neurônio são calculados e armazenados temporariamente. O \textit{backpropagation}, então, reutiliza esses valores armazenados para aplicar a regra da cadeia de forma sistemática, calculando os gradientes da última camada em direção à primeira. Isso transforma a tarefa de otimização em uma sequência de multiplicações de matrizes, permitindo que múltiplos parâmetros sejam atualizados em paralelo. A seguir, veremos um passo a passo do algoritmo de \textit{backpropagation}, inspirado na referência \cite{Marquardt2021}, de forma análoga à que usamos para atualizar os pesos do Perceptron na Seção \ref{sec:perceptron}.

\begin{enumerate}
    \item Calcule o desvio da função custo do valor real já com respeito a uma saída $l$ e sua derivada após aplicar a regra da cadeia uma vez:
    \begin{equation}
        \Delta_{l'} = (y^{[n]}_{l'} - \hat{y}_{l'})\frac{d}{dz_{l'}^{[n]}}f_{l'}(z^{[n]}_{l'}). 
    \end{equation}
    \item Agora, como as derivadas $\frac{d}{dz_{l'}^{[k]}}f_{l'}(z^{[k]}_{l'})$ foram calculadas com a etapa propagação em que calculamos $f_{l'}(z^{[k]}_{l'}) $, busque-as na memória e atualize $\Delta_{l'}$ com a composição matricial $\hat{M}$ da equação (\ref{eq:matrizcompositionbackprop}):
    \begin{equation}
        \Delta_{l'} \leftarrow \Delta_{l'}\hat{M}\frac{d}{dw^{[k]}_{il}}z^{[k]}_{l} .
    \end{equation}
    \item Repita os passos 1 e 2 para todos os $l'$'s. Em seguida, tome a média dos $\Delta_{l'}$'s e chame-a de $\Delta$. Então, atualize o peso $w_{il}^{[k]}$ como fizemos na equação (\ref{eq:SGD_w_perceptron}) e (\ref{eq:SGD_b_perceptron}):
    \begin{align}
        &w_{il,\text{novo}}^{[k]} \leftarrow w_{il,\text{antigo}}^{[k]} - \eta\Delta, \\
        &b_{l,\text{novo}}^{[k]} \leftarrow b_{l,\text{antigo}}^{[k]}- \eta\Delta. 
    \end{align}
\end{enumerate}

O algoritmo de \textit{backpropagation} apresenta limitações que podem comprometer seu desempenho e sua aplicabilidade em redes neurais profundas. Uma das principais limitações é o problema da explosão e do desaparecimento do gradiente, que ocorre quando os gradientes calculados pelo algoritmo são excessivamente grandes ou pequenos \cite{NathanKutz2017}. Este fenômeno pode resultar na amplificação ou atenuação cumulativa dos gradientes ao longo das camadas, dificultando a convergência adequada dos parâmetros da rede, pois o tamanho do passo da otimização depende da magnitude do gradiente, que, quando pequeno, resulta em convergência lenta. Isso exige muitos ciclos de treinamento, aumentando o tempo e os recursos computacionais necessários. Outra limitação importante é que os algoritmos de otimização que utilizam o \textit{backpropagation} são fortemente influenciados pela inicialização dos parâmetros. Esse fator pode resultar em desempenho insatisfatório ou na incapacidade de encontrar um ótimo local.

Além disso, o algoritmo de \textit{backpropagation} tende a ser mais eficaz quando utilizado com grandes conjuntos de dados, pois permite que a rede neural capture melhor as complexidades e variações dos padrões presentes no problema. Contudo, a exigência de grandes volumes de dados pode se tornar um obstáculo em domínios onde a coleta de dados é limitada ou envolve questões éticas e de privacidade, como em aplicações médicas ou de segurança. Nessas situações, a escassez de dados pode comprometer a capacidade da rede de aprender representações robustas \cite{GradAmplif2020}. Ademais, na ausência de técnicas apropriadas de regularização, como o \textit{dropout} ou normalização, as redes neurais podem facilmente sofrer de sobreajuste (\textit{overfitting})\cite{Bashir}. Esse fenômeno ocorre quando o modelo se ajusta excessivamente aos dados de treinamento, capturando tanto padrões relevantes quanto ruídos específicos do conjunto de dados, resultando em um desempenho insatisfatório em novos dados não vistos, devido à sua capacidade limitada de generalização.

\section{Aplicação de redes neurais na física}
\label{sec:Capitulo3}

Neste capítulo, apresentamos quatro abordagens distintas que utilizam redes neurais aplicadas ao estudo do pêndulo simples, um sistema físico clássico. Cada uma das abordagens explora diferentes aspectos do problema, empregando técnicas de ML para modelar, prever e interpretar o comportamento do sistema. Na primeira abordagem (Seção  \ref{sec:EX:Pendulo}), utilizamos aprendizado supervisionado para estimar um parâmetro físico do sistema, especificamente a aceleração da gravidade $g$, a partir de dados simulados. Em seguida, na Seção \ref{sec:EX-EDO}, aplicamos redes neurais profundas para resolver a equação diferencial ordinária que descreve o movimento do pêndulo, explorando a capacidade dessas redes de aproximar soluções complexas de equações diferenciais. Na terceira abordagem (Seção \ref{sec:EX:AutoencoderPendulo}), empregamos \textit{autoencoders} para descobrir o espaço latente do sistema, buscando reduzir a dimensionalidade dos dados e encontrar uma representação compacta e eficiente do estado do pêndulo. Por fim, na Seção \ref{sec:EX:SINDY}, utilizamos o método SINDy (\textit{Sparse Indentification of Non-linear Dynamics}), conforme desenvolvido por \cite{SINDyAutoencoder_steven2019}, visando identificar a equação diferencial subjacente que governa o sistema, utilizando imagens como entrada. Cada uma dessas abordagens ilustra o potencial das redes neurais em diferentes contextos de modelagem física, destacando a versatilidade dessas ferramentas para resolver problemas complexos eficientemente.

\subsection{Exemplo 1: Pêndulo}
\label{sec:EX:Pendulo}

Começaremos com um exemplo fundamental: o aprendizado  supervisionado de parâmetros aplicado ao pêndulo simples. Escolhemos este problema por duas razões principais. A primeira é que o modelo do pêndulo é amplamente conhecido por estudantes de graduação, tornando-o um ponto de partida acessível para introduzir conceitos de ML no contexto de sistemas físicos. A segunda razão é que o estudante já deve estar familiarizado com problemas de regressão de parâmetros, comuns em experimentos de laboratório didático, como a estimativa de constantes físicas. Neste exemplo, apresentado no Jupyter Notebook 04 \cite{Github}, visamos demonstrar uma abordagem alternativa para resolver um problema que o aluno já conhece bem, utilizando redes neurais para estimar parâmetros do sistema. Embora se trate de um  problema didático simplificado, com fins ilustrativos,  ele possibilita estabelecer a base conceitual e metodológica, preparando o terreno para as aplicações mais inovadoras e avançadas que serão discutidas nas seções subsequentes.

A equação diferencial do pêndulo é dada por: 
\begin{equation}
    \frac{d^2\theta}{dt^2} + \frac{g}{\ell}\sin{\theta} = 0,
    \label{Eq:PenduloRaw}
\end{equation}
onde $\theta$ é o ângulo em relação ao eixo vertical, $g$ é a aceleração da gravidade, e $\ell$ é o comprimento da haste do pêndulo. Na aproximação de ângulos pequenos, sabemos que ${\sin{\theta} \approx \theta}$, e logo a equação simplifica para 
\begin{equation}
    \frac{d^2\theta}{dt^2} + \frac{g}{\ell}\theta = 0,
    \label{Eq:PenduloSmallAngle}
\end{equation}
\noindent com solução conhecida,
\begin{equation}
    \theta(t) = \theta_0 \cos{\Big(\sqrt{\frac{g}{\ell}}t}\Big).
    \label{pinn-pendulo-solution}
\end{equation}

No contexto dos laboratórios didáticos, o aluno  provavelmente encontrou a expressão para o período ${T = \frac{2\pi}{\omega} = 2\pi\sqrt{\frac{\ell}{g}}}$, isolou $g$, mediu experimentalmente $T$ e $\ell$, e encontrou o valor de $g$. Neste exemplo, suponhamos que os dados experimentais, representados por $\theta(t_i)_{data}$, foram obtidos, por exemplo, por meio da filmagem do pêndulo em um dado instante de tempo $i$, ou por meio da projeção de seu movimento sobre uma fita que se desloca linearmente no tempo, como em um sismógrafo~\footnote{Isto funciona aproximadamente se $\theta$ é pequeno, pois a variação de ${y(t) = -\ell(1 - \cos{\theta}) \approx \frac{\ell}{2}\theta(t)^2 }$ é de segunda ordem.}. Suponhamos adicionalmente que medimos $\ell$ e que também sabemos $\theta_0 = \theta(0)$. 

Para esta aplicação, utilizaremos uma rede neural com um único neurônio, possuindo como entrada um vetor  $\mathbf{t}$ de tamanho $M+1$  e uma saída. Essa configuração equivale a um perceptron \sout{de entrada única}, no qual a entrada representa os instantes de tempo e a saída corresponde aos valores previstos para a aceleração da gravidade. Denotamos essa rede como ${\textit{NN}(\mathbf{t}; \mathbf{w}, \mathbf{b}) = g'}$, onde $\mathbf{t}$ representa os tempos, e os parâmetros $\mathbf{w}$ e $\mathbf{b}$ são os pesos e o \textit{bias} da rede, respectivamente.
O tempo foi particionado ao longo de um intervalo definido, representado como ${\mathbf{t} = (t_0, t_1, \ldots,t_i,\ldots, t_M)}$. Declarada a rede neural, precisamos construir nossa função custo. Para podermos comparar os dados, que são ângulos, com a saída da rede, que é uma constante $g'$. 

Portanto, utilizaremos a saída da rede na equação \eqref{pinn-pendulo-solution} antes de passar para a função custo, o que resulta em  ${\theta'(t) = \theta_0\cos{\sqrt{\frac{g'}{\ell}}t}}$. Assim, temos duas grandezas comparáveis, os dados experimentais $\theta(t_i)_{data}$, e a função customizada $\theta'(t_i)$, para cada tempo $t_i$. Essa construção fornece uma medida indireta de $g'$, e podemos então avaliar a performance da rede por meio da função custo: 
  \begin{equation}
    \mathcal{L} = \frac{1}{M}\sum_{i=0}^{M}(\theta(t_i)_{data} - \theta'(t_i))^2.
     \label{lossdata}
\end{equation}

Quando $\mathcal{L} \approx 0$, sabemos que $g \approx g'$, e logo a rede neural aproximou bem a constante desejada, e a curva gerada por $\theta'(t)$ é próxima à original. O objetivo deste exemplo é mostrar que uma rede neural pode ser usada para regressão e recuperar os parâmetros de uma curva, i.e., esta abordagem é generalizável para exemplos mais complexos, onde se tem conhecimento explícito da expressão analítica que gera a curva, mas não se têm muitos pontos. 

\begin{figure}[!ht]
    \centering
    \includegraphics[width=1\linewidth]{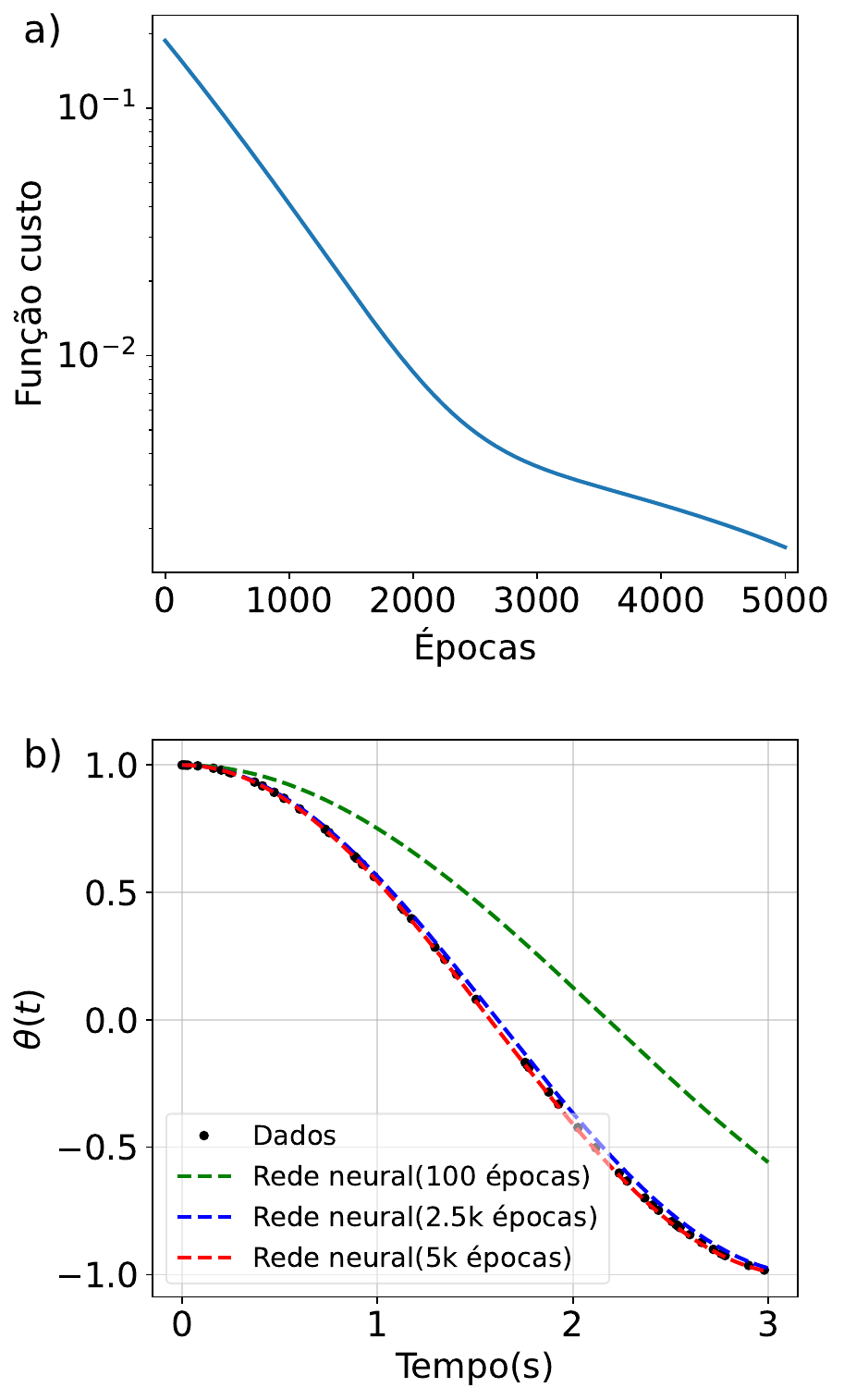}
    \caption{ Resultado do treinamento da rede neural para o aprendizado da constante gravitacional do Exemplo 1.
    (a) Evolução da função de custo ao longo das épocas de treinamento, apresentada em escala logarítmica.
    (b) Comparação entre a solução dada pela equação (24) com a constante encontrada pela rede neural em três estágios de treinamento: 100 épocas (linha verde tracejada, fase inicial), 2,5 mil épocas (linha azul tracejada, fase intermediária) e 5 mil épocas (linha vermelha tracejada, fase próxima da convergência).
    Essas escolhas visam apenas ilustrar qualitativamente a melhoria progressiva da predição à medida que o treinamento avança; quaisquer outros valores poderiam ser usados, desde que representem diferentes estágios de convergência da função de custo.}
    \label{fig:exemplo1-resultados}
\end{figure}

Na Figura~\ref{fig:exemplo1-resultados} temos dois gráficos obtidos após o treinamento da rede neural.Para esse exemplo, foi utilizada uma rede neural MLP com uma entrada (que recebe o tempo), uma camada oculta com 1 neurônio e uma saída, e função de ativação tangente hiperbólica. O treinamento consistiu em $5$ mil épocas de utilizando o otimizador \textit{Adam}, com taxa de aprendizado inicial de $10^{-2}$. Esta escolha de parâmetros foi determinada para reduzir a complexidade neste exemplo. O gráfico na Figura~\ref{fig:exemplo1-resultados}.a) representa a equação \eqref{lossdata} para cada época. Ela tem o papel de nos informar sobre o progresso do treinamento da rede neural, pois valores decrescentes da função de custo são um indicador da convergência do modelo para um resultado melhor. Porém, cabe mencionar que isto nem sempre é verdade, podem ocorrer alguns fenômenos durante a otimização, com \textit{sobreajuste} (\textit{overfitting}) no qual a rede neural se ajusta tanto aos dados de treinamento que perde a capacidade de generalizar e fazer previsões precisas sobre dados não vistos. 

Além disso, o comportamento da função custo ao longo das épocas pode fornecer pistas sobre o desempenho e a convergência do modelo. Um declínio suave e consistente na função custo ao longo das épocas sugere que o modelo está progredindo de forma estável e aprendendo de maneira eficiente. No entanto, flutuações repentinas ou uma estagnação na redução da função custo podem indicar a necessidade de ajustes nos hiperparâmetros do modelo. Portanto, é importante analisar o gráfico da função custo ao longo das épocas, sendo uma prática comum durante o ML, auxiliando na tomada de decisões para melhor otimização do modelo. Existem outros problemas que são relevantes durante o treinamento, mas que não serão tratados  de maneira mais aprofundada neste texto, como incompatibilidade entre a função custo \cite{Huang-mismatch}, platôs e/ou função custo estagnada \cite{Ainsworth}, explosão e desaparecimento/esvanecimento dos gradientes \cite{Hanin}, como também o subajuste dos modelos \cite{Bashir}.

Na Figura \ref{fig:exemplo1-resultados}.b) temos a comparação da saída da rede neural com a expressão teórica \eqref{pinn-pendulo-solution} da solução do pêndulo, mostrando um bom acordo entre as curvas. Este exemplo pode ser encontrado no \textit{Jupyter Notebook} com título `03-Exemplo 1'', nele o leitor encontrará mais  detalhes técnicos da implementação,  como também uma versão alternativa mais eficiente para resolver esse problema usando diferenciação automática.

\subsection{Exemplo 2: Resolvendo a equação diferencial}
\label{sec:EX-EDO}

Nesta seção, apresentamos como resolver equações diferenciais ordinárias (EDOs) utilizando o método de Redes Neurais Informadas com Física, do inglês \textit{Physics-Informed Neural Networks} (PINNs)~\cite{PIML-review-nature}. As redes neurais são reconhecidas por sua capacidade de aproximar funções complexas, justificando seu uso para soluções aproximadas de EDOs.
Esta aplicação é o cerne do método PINN, que integra conhecimento físico específico no processo de aprendizado da rede, como detalhado na introdução deste artigo. Para integrar a equação diferencial na aprendizagem da rede, incorporamos a equação \eqref{Eq:PenduloSmallAngle} na função custo durante o treinamento. Esta nova função é composta não apenas pela equação diferencial, mas também pelas condições iniciais e de contorno do problema, conforme expresso na seguinte equação:
\begin{equation}
   \mathcal{L} = \mathcal{L}_{\text{data}} +\mathcal{L}_{\text{EDO}} +\mathcal{L}_{\text{IC}} + \mathcal{L}_{\text{CC}}.
    \label{Loss-pinn}
\end{equation}

Aqui, $\mathcal{L}_{\text{data}}$ é utilizado quando dispomos de dados experimentais que representam parte da solução da equação, semelhante à equação \eqref{lossdata}; $\mathcal{L}_{\text{EDO}}$ reflete a contribuição da própria equação diferencial, $\mathcal{L}_{\text{IC}}$ corresponde à contribuição das condições iniciais (IC), e $\mathcal{L}_{\text{CC}}$ às condições de contorno (CC). Note que, apesar de adicionarmos restrições à função custo, neste exemplo ainda estamos tratando de aprendizado supervisionado, pois os ângulos são comparados com os ângulos preditos pela rede através do termo $\mathcal{L}_{\text{data}}$.

Para ilustrar este método, aplicaremos o PINN à equação do pêndulo simples com pequenos ângulos, resultando em um oscilador harmônico simples. A rede neural, denotada como $\textit{NN}(\mathbf{t}; \mathbf{w},\mathbf{b})$, será treinada para aproximar $\theta(\mathbf{t})$, assim a solução da equação diferencial do pêndulo que será inserida durante o treinamento tem a seguinte forma:
\begin{equation}
    \mathcal{L}_{\text{EDO}} = \sum_i \left( \frac{d^2}{dt^2}\textit{NN}(\mathbf{t}_i; \mathbf{w},\mathbf{b}) + \frac{g}{\ell}\textit{NN}(\mathbf{t}_i; \mathbf{w},\mathbf{b}) \right)^2.
\end{equation}

Como esta é a equação diferencial, seu valor mínimo ideal é zero. A condição inicial é dada apenas por $\theta_0$, sendo modelada na função custo como:
\begin{equation}
    \mathcal{L}_{\text{IC}} = \left( \textit{NN}(t=0; \mathbf{w},\mathbf{b}) - \theta_0 \right)^2.
\end{equation}

\begin{figure}[!ht]
    \centering
    \includegraphics[width=0.95\linewidth]{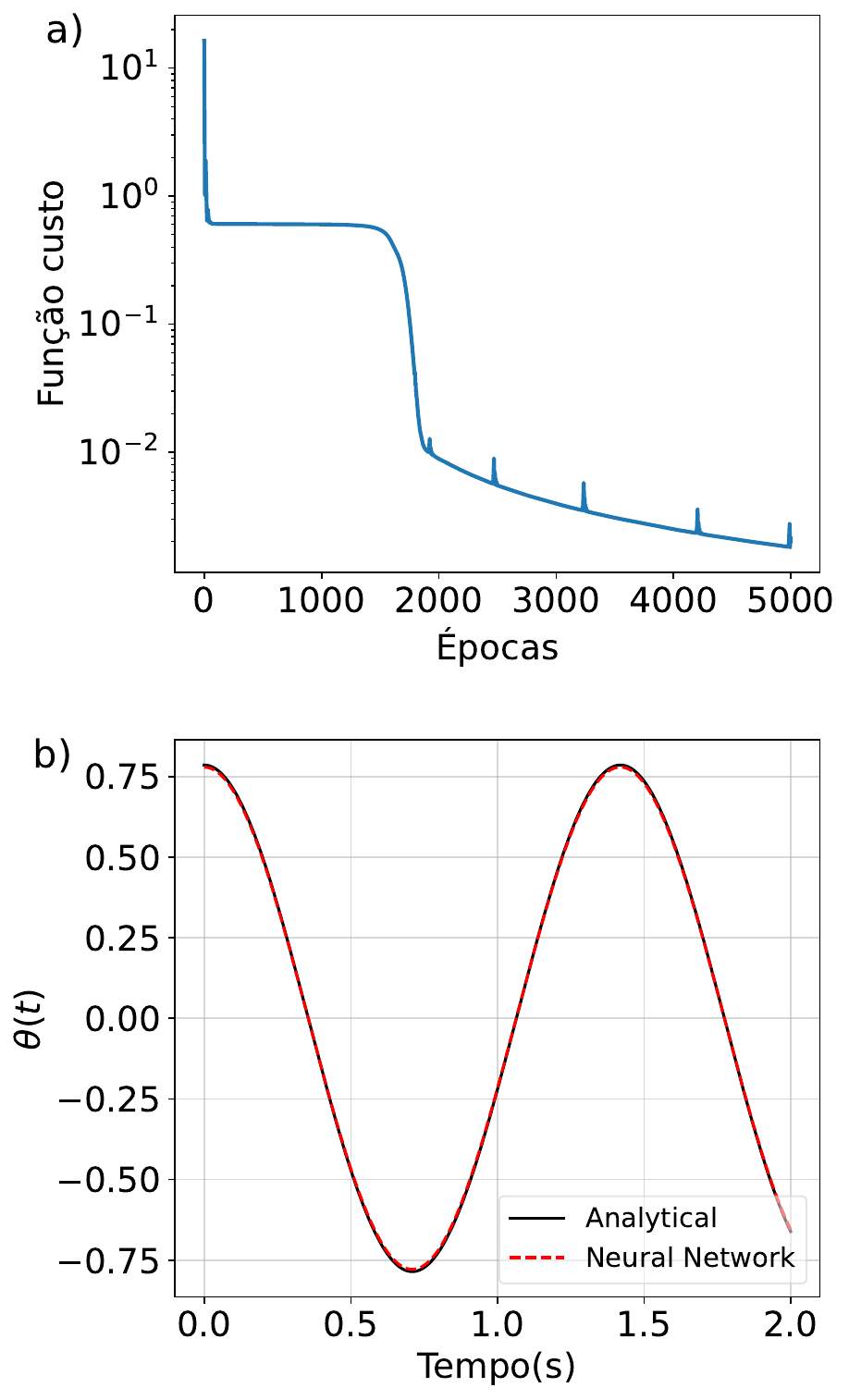}
    \caption{Resultado da resolução da equação diferencial com rede neurais  via o método PINN. Em a) o gráfico da função custo em escala logarítmica em função das épocas ilustra o processo de treinamento, que evidencia como o erro evolui para cada iteração. Em b) temos a evolução do ângulo do pêndulo simples em função do tempo obtida via solução analítica, em preto, e o resultado da rede neural após o treino, em vermelho.}
    \label{fig:pinn-edo-resultados}
\end{figure}

Com este enfoque, as PINNs permitem uma integração eficaz e direta das leis físicas na arquitetura de aprendizado profundo, abrindo caminho para soluções robustas e precisas em problemas físicos complexos.  Isso ocorre porque agora a rede neural contém conhecimento sobre o
sistema físico, e a $\textit{NN}$ é uma aproximação da solução deste modelo, considerando que satisfaz a EDO característica, por definição.  Além disso, como o aprendizado é mais estruturado (i.e., há mais informação na função custo), as fases de treino das PINNs são, em geral, mais curtas e requerem menos dados. Compare, por exemplo, os gráficos \ref{fig:exemplo1-resultados}.a) e \ref{fig:pinn-edo-resultados}.a). É notável que, apesar da função custo em (\ref{Loss-pinn}) ter mais termos positivos do que a definida em (\ref{lossdata}), o valor de $\mathcal{L}\approx 10^{-2}$ é alcançado com metade das épocas do que no caso do Exemplo~1. Para esse exemplo, foi utilizada uma rede neural MLP com uma entrada (que recebe o tempo), duas camadas ocultas com 10 neurônios cada e uma saída, com função de ativação seno. O treinamento consistiu de $5$ mil épocas, utilizando o otimizador \textit{Adam} com taxa de aprendizado de $\eta=0.01$, adicionado a um \textit{scheduler}  que reduz em $10\%$ o valor de $\eta$ a cada 250 épocas.

Na Figura \ref{fig:pinn-edo-resultados} temos dois gráficos obtidos após o treinamento da rede neural para se resolver a equação diferencial. O gráfico à esquerda (a) representa a equação \eqref{Loss-pinn} para cada época, e em (b) temos a comparação da saída da rede neural com a expressão teórica \eqref{pinn-pendulo-solution} da solução do pêndulo, mostrando um bom acordo entre as curvas. Este exemplo pode ser encontrado no \textit{Jupyter Notebook} com título ``04-Exemplo 2''. Nele o leitor encontrará os detalhes técnicos da implementação.

\subsection{Exemplo 3: Revisitando o pêndulo com \textit{Autoencoders}}
\label{sec:EX:AutoencoderPendulo}

As redes neurais também podem ser aplicadas a tarefas rotineiras em computação, trazendo novas abordagens que podem revelar propriedades inesperadas ou oferecer melhorias em termos de eficiência e desempenho. Um exemplo é o uso de redes neurais para compressão de dados, utilizando a arquitetura de codificador-decodificador, conhecida como \textit{Autoencoder}. 
O \textit{Encoder} é responsável por transformar um dado de entrada (como uma imagem) em uma representação de dimensionalidade reduzida, comprimindo a informação para um formato mais compacto (como uma lista com um número menor de bits). Essa técnica não apenas diminui o tamanho do dado, mas também pode capturar as principais características e abstrações do conjunto de dados original, o que pode ser útil para tarefas como a remoção de ruído, geração de novos dados, ou mesmo aprendizado não supervisionado de representações latentes \cite{Lusch_KoopmanAutoencoders2018}.

Considere o seguinte exemplo: uma curva definida pela função $f(t) = \sin{(\omega_1 t)} + \sin{(\omega_2 t)}$ apresenta uma expressão funcional que pode se tornar bastante complexa, especialmente quando mais frequências são adicionadas ao modelo. No entanto, apesar da curva ser composta por infinitos pontos ao longo do tempo, toda a sua informação pode ser completamente determinada conhecendo apenas duas variáveis-chave: as frequências $\omega_1$ e $\omega_2$ (aqui estamos assumindo que a fase é conhecida e $\phi_i =0$, para $i=1,2$). Isso implica que, para armazenar a curva inteira, basta saber que ela é uma soma de duas funções seno com suas respectivas frequências, representadas por dois números reais, sendo um exemplo de como uma descrição paramétrica pode ser extremamente compacta. 

De forma semelhante, os fractais exemplificam outro caso de complexidade emergente a partir de regras simples: figuras altamente complexas e visualmente ricas podem ser geradas por meio de equações iterativas relativamente simples. Esses exemplos ilustram como informações complexas podem ser compactamente representadas por um conjunto reduzido de parâmetros ou instruções \cite{Gamelin2001}.

Outro exemplo de aplicação de \textit{encoder} é na compressão de imagens em preto e branco, cuja representação dos píxeis é dada por uma matriz bidimensional (2D). Quando uma imagem é submetida à transformada discreta de Fourier, a informação espacial dos píxeis é convertida em uma representação no domínio da frequência. Nesse processo, a maioria da informação significativa da imagem pode ser capturada por um conjunto relativamente pequeno de coeficientes de frequência (isto é conhecido como a aplicação de um filtro {\it passa baixa}, ou em inglês {\it low pass filter} \cite{NathanKutz2017}). Assim, ao invés de armazenar todos os valores dos píxeis, é possível representar a imagem inteira de maneira compacta, utilizando apenas as frequências mais relevantes que compõem sua estrutura.

Um algoritmo conhecido que comprime dados por meio da Transformada Discreta do Cosseno (DCT) é o formato de imagens JPEG \cite{NathanKutz2017}. Fornecido um dado comprimido, para acessar a versão original, é necessário haver um algoritmo que realize a operação inversa, que, tomando o dado comprimido, recupere (mesmo que parcialmente) a imagem inicial. O algoritmo que realiza essa tarefa é chamado de \textit{Decoder}, e, no caso ilustrado acima, seria a transformada inversa de Fourier.

Aplicando-se ambos algoritmos concatenados, \textit{encoder} e depois \textit{decoder}, temos um algoritmo resultante chamado de $\textit{Autoencoder}$. O nome \textit{auto} sinaliza que uma parte do algoritmo é a inversa do outro, então a entrada e saída de um \textit{autoencoder} devem ser correspondentes. Em geral, essa correspondência não é perfeita, pois sempre há perda de informação em um processo de compressão,  e o uso da palavra `inversa' é, na verdade, um abuso de notação, porém, um bom \textit{autoencoder} deve ser capaz de recuperar a imagem original com boa fidelidade. 

\begin{figure}[!ht]
    \centering
    \includegraphics[scale=0.35]{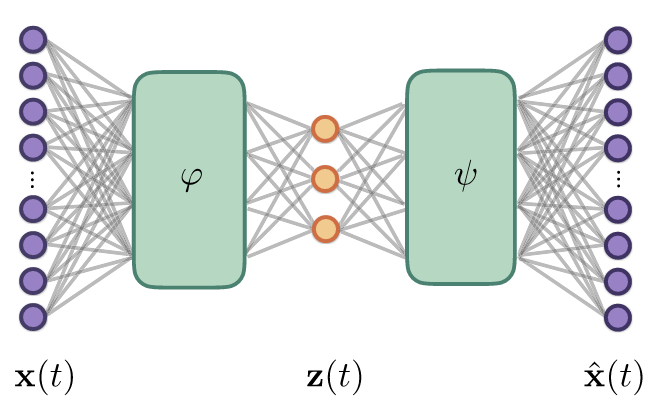}
    \caption{Representação pictórica de uma rede neural \textit{Autoencoder}.  A entrada é representada por $\mathbf{x}(t)$, passa por uma camada de neurônios (geralmente CNN), e então é comprimida para um número menor de neurônios pelo \textit{encoder} $\varphi$, de forma que $\varphi(\mathbf{x(t)})=\mathbf{z}(t)$. O \textit{decoder}, a saída, $\psi$ recupera a entrada original (aproximadamente) com a operação $\psi(\mathbf{z}(t)) = \hat{\mathbf{x}}(t)$. O objetivo do \textit{autoencoder} após treinado é reproduzir a relação $\psi\circ\varphi = \mathbb{I}$. Aqui $\varphi$ é composto por redes CNN seguidas de uma rede neural totalmente conectada, e $\psi$, o mesmo, porém com a ordem das camadas invertida. Fonte \cite{SINDyAutoencoder_steven2019}. }
    \label{fig:autoencoder}
\end{figure}

No caso das redes neurais, podemos construir uma arquitetura para um \textit{autoencoder}, como ilustrado na Figura \ref{fig:autoencoder}. Temos uma entrada $\mathbf{x}(t)$ que fornece informações para um \textit{encoder} $\varphi$, cuja saída é uma camada oculta com menos neurônios, $\mathbf{z}(t)$. Neste exemplo, e nos que seguem, assumimos que $\varphi$ é composto por uma série de camadas convolucionais (CNN) seguidas de camadas totalmente conectadas. Assim, como na compressão de imagens, $\varphi$ reduz a entrada $\mathbf{x}(t)$, que contém muitas informações, para alguns poucos números em $\mathbf{z}(t)$, chamados de \textit{espaço latente} ou \textit{camada latente} (às vezes também chamada de {\it code}). A tarefa do operador $\psi$ é então decodificar a informação comprimida e recuperar a entrada inicial de forma aproximada: $\psi(\mathbf{z}(t)) = \hat{\mathbf{x}}(t) \approx \mathbf{x}(t)$. É prática comum que a rede de $\psi$ seja a ordenação contrária de $\varphi$, i.e., camadas totalmente conectadas seguidas de camadas CNN. Então, a tarefa do \textit{autoencoder} é aproximar a relação \footnote{ Note que $\psi\circ\varphi$ representa a composição de $\psi$ com $\varphi$, e não o produto interno. } 

$$\psi \circ \varphi \approx \mathbb{I},$$ 
onde $\mathbb{I}$ é o operador identidade no espaço vetorial que contém $\mathbf{x}(t)$, i.e.,  queremos que $\psi (\phi (\mathbf{x}(t))) \approx \mathbf{x}(t)$.  Assim, quando declararmos a função de custo do tipo MSE para a otimização, teremos 
\begin{equation}
    \mathcal{L}_{\text{data}} = ||\mathbf{x}(t) - \hat{\mathbf{x}}(t)||^2_2 =|| \mathbf{x} - \psi \circ \varphi (\mathbf{x})||^2_2.
\end{equation}

Este processo representa uma mudança fundamental em relação ao que fizemos anteriormente. Em vez de comparar a saída da rede com um rótulo externo, a otimização ocorre ao comparar a saída $\mathbf{z}(t)$ com a própria entrada. Por não haver uma supervisão explícita, este método é um exemplo de aprendizado não supervisionado.

Didaticamente, considere o seguinte cenário: desejamos aplicar um \textit{autoencoder} em uma situação onde o leitor pretende enviar uma imagem para outra pessoa, mas possui a limitação de transmitir apenas 3 números reais. Se tivermos uma rede treinada, podemos ``cortar'' o \textit{autoencoder} ao meio, mantendo o \textit{encoder} e enviando o \textit{decoder} ao destinatário da imagem. A imagem original é comprimida pelo \textit{encoder} $\varphi$, gerando $\mathbf{z}(t)$. Assim, podemos enviar $\mathbf{z}(t)$ para a outra ponta do canal de comunicação, onde será decodificada com $\psi$, recuperando a mensagem original. Em física, temos outros interesses. Por exemplo, no reconhecimento de imagens, essas arquiteturas podem ser usadas para limpeza de dados originais. Em técnicas observacionais em astrofísica, \textit{autoencoders} podem remover o ruído de ondas eletromagnéticas, facilitando a visualização e o processamento dos dados \cite{Gheller_2021}.

No exemplo que segue, iremos revisitar o pêndulo com \textit{autoencoders}  e analisaremos o comportamento do espaço latente. Vamos supor que nossos dados sejam uma sequência de \textit{frames} originados de um vídeo da dinâmica de um pêndulo com apenas uma oscilação, partindo de um ângulo inicial $\theta_0=\theta(t=0)$ qualquer e velocidade inicial igual a zero, $\dot{\theta}(0)=0$. Para gerar dados análogos a um vídeo da dinâmica do pêndulo, nosso conjunto de dados é composto por uma função que toma o resultado numérico da equação (\ref{Eq:PenduloRaw}) do pêndulo, $\theta(t)$, e o converte em uma imagem de duas dimensões $\mathbf{x}(t)$. 

O objetivo deste exercício é verificar se o espaço latente do \textit{autoencoder} com as imagens do pêndulo pode ser reduzido a apenas uma variável. Isto deveria ser possível, dado que a equação do pêndulo é um sistema que pode ser descrito por uma única variável $\theta(t)$. Esperamos, então, que um \textit{autoencoder} com espaço latente ($\mathbf{z}(t)$), de dimensão um, deva ser capaz de reproduzir as imagens originais, mesmo no regime não linear de ângulos grandes da equação do pêndulo. Veja que a não linearidade das redes neurais nos permite exigir que a compressão ocorra mesmo para a equação diferencial de grandes ângulos, o que não é possível com técnicas lineares de redução de dimensionalidade, como \textit{Principal Component Analysis} (PCA) \cite{SINDyAutoencoder_steven2019}, pois estas exigem que haja uma linearização em algum momento da compressão, e isto pode levar a perdas na representação.

\begin{figure}[!ht]
    \centering
    \includegraphics[width=1\linewidth]{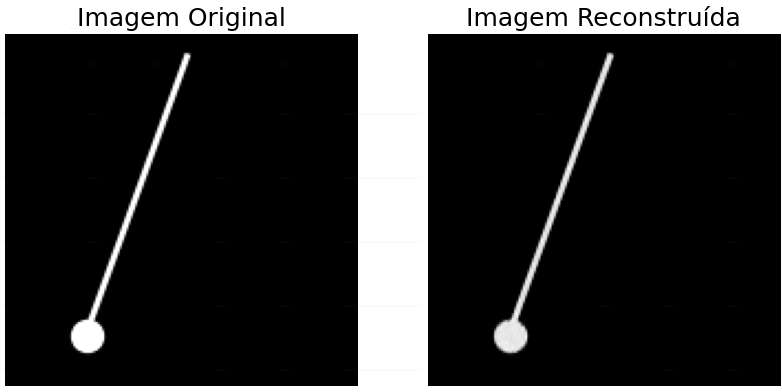}
    \caption{Comparação entre a imagem original e a reconstruída utilizando um \textit{autoencoder} com espaço latente igual a 1. A imagem à esquerda representa a imagem original e a imagem à direita é a versão reconstruída da mesma imagem.}
    \label{fig:imagem_reconstruida}
\end{figure}

Para realizar o treinamento foi utilizada uma \textit{autoencoder} com 2 camadas convolucionais, a primeira com 5 filtros e a segunda com 10 filtros, seguidas de uma camada de \textit{pooling} com kernel de tamanho $K=2$ e 3 camadas densas para o \textit{encoder}. Similarmente, o \textit{decoder} tem as mesmas camadas, mas em ordem inversa. As camadas densas do \textit{encoder} contêm $1.6\times 10^4$, $1600$ e $40$ neurônios, respectivamente. O tamanho da saída da última camada depende do valor do espaço latente; na primeira abordagem, usamos 1, depois 2 neurônios. Para todos os neurônios, a função de ativação é do tipo ReLU. O treinamento foi composto por $2$ mil épocas, e o otimizador \textit{Adam} com taxa de aprendizado $\eta=0.001$, cujo valor foi reduzido em $10\%$ a cada 500 épocas.
O resultado do nosso exemplo pode ser visto na Figura \ref{fig:imagem_reconstruida}, que apresenta a comparação entre a imagem original (à esquerda) e a reconstruída (à direita) após o treinamento \footnote{Neste exemplo, é recomendado que o leitor utilize processamento em GPU (por exemplo, disponível gratuitamente no Colab da Google), pois o tamanho da rede utilizada torna a fase de treinamento longa.}. 

\begin{figure}[!ht]
    \centering
    \includegraphics[width=1\linewidth]{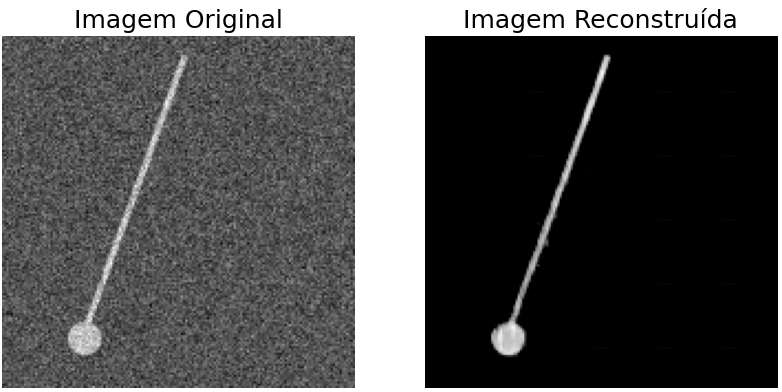}
    \caption{Comparação entre a imagem original com ruído adicionado e a imagem reconstruída utilizando um \textit{autoencoder} com espaço latente igual a 2. A imagem à esquerda representa a imagem original com ruído adicionado e a imagem à direita é a versão reconstruída da mesma imagem. Nesta imagem o ruído é de $20\%$. No material suplementar, notamos que a rede ainda recupera bem o objeto e posicionamento com ruído de até $40\%$.}
    \label{fig:imagem_reconstruida_wnoise}
\end{figure}

CNNs são notórias em abstrair somente informações essenciais dos dados, no sentido de que suas representações podem ser tais que ignoram entradas descorrelacionadas nos dados \cite{Vincent2008}. Assim, podemos tomar um modelo treinado somente com imagens do pêndulo, sem ruído, e expor o modelo a dados ruidosos (onde introduzimos um ruído de distribuição homogênea, logo descorrelacionada), como na Figura \ref{fig:imagem_reconstruida_wnoise}. Veja que, apesar das novas imagens serem perturbadas por um ruído de $20\%$, i.e., com probabilidade $P=1/5$ de substituir um píxel por um valor aleatório, ainda assim o modelo consegue reproduzir o comportamento esperado e sem o ruído. Com base nos modelos treinados, verificou-se que, ao aumentar a dimensão do espaço latente para dimensão 2, o ruído poderia atingir até $40\%$ antes de a rede apresentar sinais de baixa performance. Esta propriedade é bem conhecida na área, usada, por exemplo, em astrofísica observacional para limpeza de imagens \cite{Geller2022}. Os testes discutidos acima podem ser encontrados no material complementar, i.e., no Notebook ``06-Exemplo 3''. 

\begin{figure}[!ht]
    \centering
    \includegraphics[width=1\linewidth]{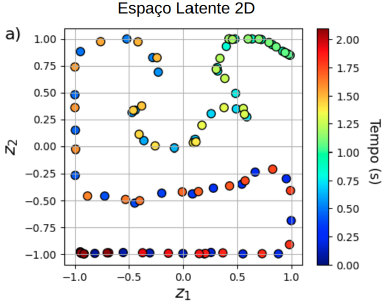}    
    \includegraphics[width=1\linewidth]{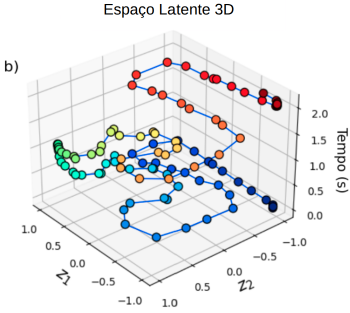}
    \caption{Visualização em 2D em a) e 3D em b) do espaço latente composto por z$_1$ e z$_2$, em função do tempo. Em ambas as imagens, cada ponto representa a posição da imagem do pêndulo no espaço latente em um instante de tempo $t_i$.}
    \label{fig:espaço-latente}
\end{figure}

Podemos agora analisar como cada \textit{frame} se distribui no espaço latente, i.e., como está representado cada segundo no espaço codificado pela rede. Para ilustrar, treinamos a rede em um espaço latente de dimensão dois, $z_1(t)$ e $z_2(t)$, veja a Figura \ref{fig:espaço-latente}. Na imagem (a), temos um gráfico em 2D, com o eixo horizontal representando $z_1$ e o eixo vertical $z_2$. Cada círculo representa o espaço latente para um \textit{frame} do vídeo, que corresponde a cada instante de tempo do pêndulo. 

Para identificar qual o instante de tempo do \textit{frame} em cada círculo, utilizamos uma cor, variando de 0 (em azul) até $2.1$ segundos (em vermelho), explicitadas no mapa de cores à direita da imagem. Note que, ao longo da curva, não há uma separação contínua nos instantes de tempo do pêndulo na figura à direita, i.e., pelo mapa de cores, temos as cores azuis e vermelhas intercaladas, enquanto estas deveriam estar idealmente em pontas opostas da curva contínua que gera estes pontos: este é o caráter não linear da codificação. 
Este resultado se deve ao fato de que, como a posição inicial e final do pêndulo após um período $T$ devem ser muito próximas, a codificação deve mapear estes dados próximos, já que são quase a mesma imagem. Porém, apesar da rede ter mapeado os trechos de tempo $[0,T/2]$ e $[T/2,T]$ de maneira sobreposta, com ordem contrária, ainda é possível observar que a codificação está inteiramente contida em uma curva (i.e., uma superfície de dimensão 1). 
Para garantir a continuidade da curva que contém a codificação, podemos adicionar mais um eixo, em que ordenamos os dados na mesma ordem em que os \textit{frames} são apresentados, o resultado pode ser visto na Figura \ref{fig:espaço-latente} (b) onde temos o mesmo gráfico em 3D. Assim, podemos visualizar a evolução do espaço latente em função do tempo, e observa-se que a rede neural foi capaz de codificar corretamente a posição do pêndulo em uma superfície de dimensão 1 parametrizada pelo tempo. 

Esta abordagem mostra que podemos alavancar a capacidade de representação dos \textit{Autoencoders} para estimar a dimensionalidade de um sistema. Neste tipo de aplicação, não nos interessa a complexidade da codificação resultante, mas somente a dimensão mínima em que conseguimos codificar o sistema, pois isso nos indica quantas coordenadas são necessárias para descrevê-lo.

\subsection{Exemplo 4: \textit{Autoencoders} e arquiteturas SINDy para descoberta de EDOs }
\label{sec:EX:SINDY}
Como discutimos na introdução, há boas justificativas para defender uma posição cética com relação a redes neurais, pois, por vezes, seu funcionamento pode ser mais opaco do que o próprio sistema que estudamos. Afinal, uma rede neural pode ter milhares (ou trilhões, em casos recentes) de parâmetros para serem ajustados que atuam de maneira não linear, sobrepujando a capacidade humana de análise \cite{MSchuld2021}. Dito isto, há um esforço na comunidade para aliar conhecimentos prévios sobre sistemas físicos para reduzir a opacidade das redes neurais e podermos obter resultados que forneçam uma perspectiva mais detalhada sobre como uma rede neural opera, sem perda de generalidade. 

Neste último exemplo, aplicaremos os conceitos desenvolvidos nos exemplos anteriores para explorar uma arquitetura recente e de especial interesse para a comunidade científica. O foco será a Identificação Esparsa de Dinâmica Não Linear (SINDy, \textit{Sparse Identification of Nonlinear Dynamics}) \cite{SINDyAutoencoder_steven2019}. O método SINDy busca identificar a dinâmica subjacente de um sistema não linear, restringindo o espaço latente da rede neural de modo a revelar suas equações de movimento, enquanto a rede, como um todo, mantém a funcionalidade de um \textit{autoencoder}.

Suponha então que temos um sistema não linear, neste caso gerado pela equação (\ref{Eq:PenduloRaw}), e que produza um conjunto de dados de imagens $\mathbf{x}(t)$. Por exemplo, uma filmagem de um pêndulo com ângulos grandes. Suponha também que o espaço latente possa ser representado por uma equação diferencial (possivelmente não linear) como
\begin{equation} \label{eq:latent_space_eq}
    \frac{d}{dt}\mathbf{z}(t) = \mathbf{g}(\mathbf{z}(t)).
\end{equation} 

\noindent ou
\begin{equation} \label{eq:latent_space_eq_2nd}
    \frac{d^2}{dt^2}\mathbf{z}(t) = \mathbf{g}(\mathbf{z}(t)).
\end{equation} 

A seguir, utilizamos como exemplo canônico do SINDy o caso em que $\mathbf{z}(t)$ satisfaz uma equação diferencial de primeira ordem no que segue, porém, também usaremos a segunda ordem para o exemplo particular do sistema do pêndulo não-linear. Em princípio, poderíamos ter escolhido qualquer ordem. A seguir, apresentaremos a construção de modelos para a primeira e segunda ordem em paralelo, destacando que as mesmas regras de derivação podem ser aplicadas iterativamente para alcançar ordens superiores. No material suplementar em Jupyter notebooks, exploramos explicitamente (e somente) o caso da segunda derivada a fim de recuperar a dinâmica do pêndulo. 

Se \textit{autoencoders} são capazes de recuperar as imagens originais com uma só variável em $\mathbf{z}(t)$ (no espaço latente), que podemos assumir ser uma transformação afim de $\theta(t)$ (a função geradora das imagens), que é solução da equação (\ref{Eq:PenduloRaw}), então podemos exigir que também respeite a mesma equação diferencial, i.e., que  seja solução de 
\begin{equation}
    \ddot{z}(t) + \sin{(z(t))} =0.
    \label{eq:idealEqoutput}
\end{equation} 
Por enquanto, suponha também que desconhecemos a equação do pêndulo, i.e., nosso conjunto de dados é um conjunto de imagens do tipo $\{\mathbf{x}(t), \dot{\mathbf{x}}(t), \ddot{\mathbf{x}}(t)\}$ (lembre que podemos tomar a derivada de uma imagem numericamente, sem conhecer sua expressão analítica, desde que $\mathbf{x}(t)$ seja suficientemente suave). Trataremos de responder à seguinte pergunta no que segue: como podemos garantir que a equação para $z$ represente de fato nosso sistema? 

Para isso, criamos uma biblioteca de $m$  funções possíveis para o sistema $\{z, z^2, \ldots, z^n, \sin(z), \cos(z), \sqrt{z}, 1/z, \ldots\}$. Cada uma destas é uma entrada de uma matriz de funções $\boldsymbol{\Theta}$ (aqui,  com tamanho $1\times m$). Também vamos definir uma matriz ($m\times 1$) de coeficientes $\boldsymbol{\Xi} = (\xi_1, \ldots, \xi_m)$, de forma que a equação (\ref{eq:latent_space_eq}) seja reescrita como 
\begin{equation} \label{eq:latent_space_SINDy}
    \dot{z}(t) = \boldsymbol{\Theta}(z(t))\boldsymbol{\Xi}.
\end{equation}

\begin{figure} [h]
    \centering
    \includegraphics[scale=0.36]{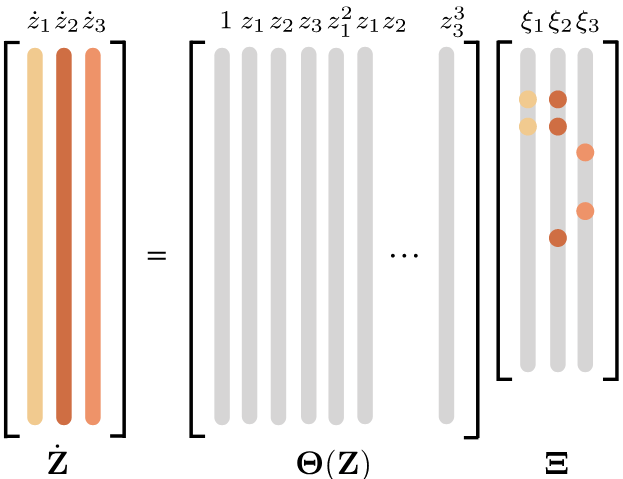}
    \caption{Generalização da equação (\ref{eq:latent_space_SINDy}) para mais variáveis. O autoencoder SINDy realiza uma busca entre diversas funções de uma biblioteca $\boldsymbol{\Theta}$, declarando a derivada do espaço latente $\mathbf{z}$ como uma combinação linear da biblioteca com coeficientes dados por $\boldsymbol{\Xi}$. Esta equação é usada como parte da função custo para a otimização, que, iterativamente, suprime alguns coeficientes $\xi_{ij}$, até que $\boldsymbol{\Xi}$ esteja suficientemente esparsa, e a equação nos retorne um modelo simples. Uma otimização bem sucedida garante que o espaço latente nos dê uma expressão analítica para as equações de movimento, e, ao mesmo tempo, contendo poucos termos, graças a esparsidade do sistema encontrado. Fonte \cite{SINDyAutoencoder_steven2019}
    .}
    \label{fig:SINDy_latent_space_Model}
\end{figure}

Para um espaço latente maior, veja a representação gráfica desta equação na Figura \ref{fig:SINDy_latent_space_Model}.  No caso de segunda ordem, no lugar da equação (\ref{eq:latent_space_SINDy}), tomamos 
\begin{equation} \label{eq:latent_space_SINDy_2nd}
    \ddot{z}(t) = \boldsymbol{\Theta}(z(t))\boldsymbol{\Xi}.
\end{equation}

\noindent Note que a equação (\ref{eq:latent_space_SINDy_2nd}) não é a segunda derivada da equação (\ref{eq:latent_space_SINDy}), mas sim uma equação alternativa no caso em que tratamos sistemas caracterizados por equações diferenciais de segunda ordem.

Para calcular a derivada de $\dot{z}$ em relação ao tempo, usamos a expressão $\dot{\mathbf{x}}(t)$ através da regra da cadeia, da mesma forma que fizemos para o algoritmo de \textit{backpropagation} na Seção \ref{sec:RNProfunda}. Neste caso, tomando a notação da Figura \ref{fig:autoencoder}, temos que $z = \varphi(\mathbf{x})$, logo
\begin{equation} \label{eq:zdot_SINDY}
    \dot{z} = \nabla_x \varphi (\mathbf{x}) \dot{\mathbf{x}}.
\end{equation}
No caso paralelo segunda derivada é tomada da mesma maneira, veja o material suplementar em \cite{SINDyAutoencoder_steven2019}, onde obtemos
\begin{equation} \label{eq:SINDy_2ndDeriv}
    \ddot{z} = \nabla_x^2 \varphi (\mathbf{x}) \dot{\mathbf{x}}^2 + \nabla_x \varphi (\mathbf{x}) \ddot{\mathbf{x}}.
\end{equation}

De forma análoga, podemos impor que a derivada da nova imagem gerada, $\dot{\tilde{\mathbf{x}}}(t)$, reproduza fielmente as derivadas das imagens do conjunto de dados, $\dot{\mathbf{x}}(t)$. Essa restrição, aplicada por meio da minimização de uma função de custo definida a seguir, assegura que $z(t)$ seja um modelo representativo do sistema, assim como o modelo original em $t$. No nosso caso, imporemos essa condição à segunda derivada, ou seja, $\ddot{\tilde{\mathbf{x}}} \approx \ddot{\mathbf{x}}$.

Ademais, para que $z$ seja também um modelo das imagens geradas, iremos exigir que a relação $$\frac{d^n}{dz^n}\tilde{\mathbf{x}}(z) \approx \frac{d^n}{dt^n}\mathbf{x}(t),$$ seja verdadeira, e então que $z$ se comporte como $\theta$, não apenas para o conjunto de dados mas também para os dados gerados. Assim, ainda seguindo a notação da Figura \ref{fig:autoencoder}, onde ${\psi(z) \approx \tilde{\mathbf{x}}}$, temos que
\begin{equation} 
    \dot{\mathbf{x}} \approx \frac{d}{dz}\tilde{\mathbf{x}} = \nabla_z \psi (z) \boldsymbol{\Theta}(z) \boldsymbol{\Xi}.
\end{equation}

\noindent Onde substituímos $\dot{z}$ pela equação (\ref{eq:latent_space_SINDy}). Para segunda ordem temos
\begin{equation} 
    \ddot{\mathbf{x}} \approx \frac{d^2}{dz^2}\tilde{\mathbf{x}} = \nabla^2_z \psi (z) \dot{z}^2 +  \nabla_z \psi (z)\boldsymbol{\Theta}(z) \boldsymbol{\Xi}.
\end{equation}
Desta vez, substituímos $\ddot{z}$ pela equação (\ref{eq:latent_space_SINDy_2nd}).

Dessa maneira, para o caso da derivada de primeira ordem, temos três exigências para nossa rede, o que se traduz em três termos para nossa função custo: Primeiro, o termo que tínhamos no Exemplo \ref{sec:EX:AutoencoderPendulo}, de reconstrução da imagem original
\begin{equation} 
    \mathcal{L}_{\text{recon}} = || \mathbf{x} - \psi \circ \varphi (\mathbf{x})||^2_2,
    \label{eq:SINDy_reconloss}
\end{equation}

\noindent adicionado ao segundo termo, de reconstrução da derivada da imagem em termos de $z$
\begin{equation} 
    \mathcal{L}_{\dot{\mathbf{x}}} = ||\dot{\mathbf{x}} - \nabla_z \psi (z) \boldsymbol{\Theta}(z) \boldsymbol{\Xi} ||^2_2,
    \label{eq:SINDy_dxloss}
\end{equation}

\noindent e finalmente o termo relativo à derivada do modelo gerado por $z$
\begin{equation}
\label{eq:SINDy_dzloss}
    \mathcal{L}_{\dot{z}} =  || \nabla_x \varphi(\mathbf{x})\dot{x} -  \boldsymbol{\Theta}(z(t))\boldsymbol{\Xi} ||^2_2.
\end{equation}

\noindent No caso do modelo para sistemas de equação com segunda derivada, 
reescrevemos as duas funções de custo acima para o caso de segunda
 ordem como
\begin{equation} \label{eq:SINDy_dxloss_2nd}
    \mathcal{L}_{\ddot{\mathbf{x}}} = ||\ddot{\mathbf{x}} - (\nabla^2_z \psi (z) \dot{z}^2 +  \nabla_z \psi (z)\boldsymbol{\Theta}(z) \boldsymbol{\Xi}) ||^2_2,
\end{equation}

\noindent e finalmente o termo relativo à derivada do modelo gerado por $z$
\begin{equation}
\label{eq:SINDy_dzloss_2nd}
    \mathcal{L}_{\ddot{z}} =  ||\nabla_x^2 \varphi (\mathbf{x}) \dot{\mathbf{x}}^2 + \nabla_x \varphi (\mathbf{x}) \ddot{\mathbf{x}} - \boldsymbol{\Theta}(z) \boldsymbol{\Xi}||^2_2.
\end{equation}

É importante notar que o cálculo destas funções é eficiente no seguinte sentido: assim como para \textit{backpropagation}, podemos guardar os valores das derivadas $\nabla \varphi$ e $\nabla \psi$, ao calcularmos $\tilde{\mathbf{x}}$, e usá-las ambas para o cálculo de $\mathcal{L}$ e para a atualização dos pesos e \textit{bias}.

Com os termos definidos acima, temos o que é necessário para encontrar modelos analíticos para a equação de dinâmica em $z$, porém, ainda não explicamos como o modelo consegue ser esparso, como sugere o título da arquitetura. Veja que, se otimizarmos a rede como está até agora, obteremos modelos do tipo $\ddot{z} = \xi_1 z + \xi_2 z^2 \ldots \xi_{n+1} \sin(z) + \ldots$. Este não é um modelo particularmente informativo sobre o sistema, já que tem tantos termos quanto forem dados. Duas novas restrições são então necessárias: primeiro, vamos exigir que $\boldsymbol{\Xi}$ tenha uma norma $||.||_1$ pequena, forçando que os valores de $\xi_i$ estejam próximos do intervalo $(-1,1)$. Adicionalmente, após alguns passos da otimização, os valores de $\xi_i$ estarão em uma distribuição tal que alguns destes serão maiores que outros (em valor absoluto), simplesmente por serem mais relevantes para a dinâmica do que outros, assim adicionemos um corte na otimização que elimine os coeficientes $|\xi_i| <0.1 $, tal que $\xi_i=0$, após o corte. Com isso, estamos eliminando elementos da matriz $\boldsymbol{\Xi}$ e tornando-a esparsa, forçando nosso modelo a ser o mais simples possível.  

\begin{figure}[ht]
     \centering
     \includegraphics[scale=0.36]{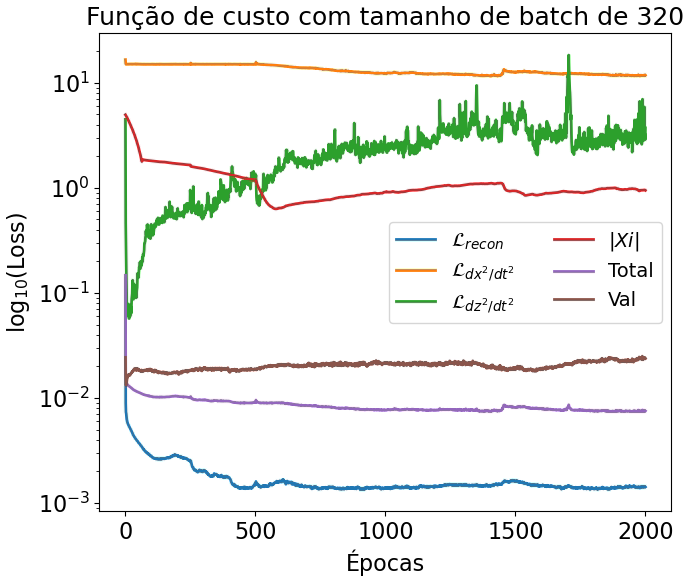}
     \caption{Múltiplas funções de custo do modelo de \textit{Autoencoders} SINDy em função das épocas de treinamento. No eixo $y$ temos na escala logarítmica a métrica de erro  e no eixo $x$ temos o número de épocas. A curva azul representa os valores da função de custo parcial na equação (\ref{eq:SINDy_reconloss}), em laranja, a equação (\ref{eq:SINDy_dxloss_2nd}), em verde, a equação (\ref{eq:SINDy_dzloss_2nd}), e, em vermelho, o termo $||\boldsymbol{\Xi}||_1$ da equação (\ref{eq:SINDy_lossfunction}). A curva correspondente à função de custo total (Total, roxa), dada pela equação (\ref{eq:SINDy_lossfunction}), apresenta uma tendência decrescente, refletindo a otimização global do modelo. Em algumas iterações do processo de treinamento, encontramos picos na função de custo, isso é representativo do estágio de remoção de coeficientes pequenos do modelo, $|\xi_i| <0.1 $, para torná-lo esparço, neste momento a performance da rede é prejudicada temporariamente. A curva marrom (Val) é dada pelo cálculo da função de loss relativa aos dados de validação, dados nunca vistos pela rede, para avaliar sua performance fora do conjunto de dados.}
     \label{fig:SINDy_loss}
\end{figure}

Nossa função custo toma, então, a forma final 
\begin{equation} \label{eq:SINDy_lossfunction}
    \mathcal{L}_{\text{SINDy}} = \mathcal{L}_{\text{recon}} + \lambda_1\mathcal{L}_{\dot{x}} + \lambda_2\mathcal{L}_{\dot{z}} + \lambda_3||\boldsymbol{\Xi}||_1,
\end{equation}
\noindent onde $\lambda_i$ são hiperparâmetros a serem ajustados, e para o caso do pêndulo, foram usados os valores de $ {\lambda_1 =5 \times 10^{-4}, \lambda_2 =5 \times 10^{-5}, \lambda_3 = 10^{-5}}$, os mesmos usados no material suplementar em \cite{SINDyAutoencoder_steven2019}.  O gráfico com os valores parciais da função custo pode ser encontrado na Figura \ref{fig:SINDy_loss}. O treinamento foi realizado com uma rede neural MLP contendo 5 camadas ocultas para o \textit{enconder} e outras 5 para o \textit{decoder}, com $256,128,64$ e $32$ neurônios nas camadas, conectados por uma camada latente com $1$ neurônio. Como função de ativação foi utilizada a função ReLU, $2$ mil épocas de treino, e otimizador \textit{Adam} com taxa de aprendizado de $10^{-4}$. 

Note que há uma diferença fundamental entre duas possíveis abordagens de treinamento da rede. No nosso caso, a rede é treinada desde o início com uma função de custo que impõe a necessidade de encontrar e respeitar uma equação diferencial simultaneamente. 
Na segunda abordagem, a rede é primeiramente treinada para reconhecimento de imagens e, posteriormente, o espaço latente é ajustado de forma independente. 
Neste outro caso, ao otimizar a rede neural em duas etapas distintas, cada uma resulta em um espaço de parâmetros diferente. Consequentemente, a primeira otimização pode conduzir a um mínimo local distante das soluções ótimas para a segunda. 
No caso das arquiteturas SINDy, como ambas as condições são otimizadas simultaneamente, a rede neural busca um caminho no espaço de parâmetros que satisfaça todas as restrições simultaneamente.
A prioridade atribuída a cada parâmetro é, de certa forma, determinada pelos hiperparâmetros $\lambda_i$. Assim como ocorre com outros hiperparâmetros, a escolha adequada desses valores exige familiaridade e experiência com a técnica \cite{carleo_machine_2019}.  Quanto ao caráter do aprendizado, em nenhum momento comparamos diretamente a equação encontrada (
ef{eq:idealEqoutput}) com a equação real do sistema (
ef{Eq:PenduloRaw}), i.e., não informamos à rede qual é a equação que deveria ter encontrado, o que torna este exemplo um caso de aprendizado não supervisionado. 

A arquitetura SINDy pode ser usada de maneira mais generalizada para sistemas não lineares de mais variáveis, como atratores de Lorenz e reações de difusão \cite{SINDyAutoencoder_steven2019}. Neste caso, a equação (\ref{eq:latent_space_SINDy}) se torna um sistema matricial como da Figura \ref{fig:SINDy_latent_space_Model}. Lembre-se que, neste caso, precisamos também tomar os polinômios cruzados como ${\{z_1^{k_1}z_2^{k_2}\ldots z_q^{k_q} | k_1,k_2,\ldots, k_q \in \{1,2,3, \ldots, n\}\}}$, onde $k_q$ é a ordem do monômio correspondente à dimensão $q$, na biblioteca $\boldsymbol{\Theta}$, para reproduzir sistemas de equações acopladas.

No material complementar deste artigo, inspirado em \cite{GithubSINDy}, apresentamos o exemplo trabalhado para a equação não linear do pêndulo, em que nosso conjunto de dados é composto por imagens do tipo  $\{\mathbf{x}(t), \dot{\mathbf{x}}(t), \ddot{\mathbf{x}}(t)\}$. Aqui, restringimos a energia do pêndulo para o caso em que ${|\dot{\theta}^2(0)/2 - \cos{(\theta(0))}|\leq 0.99}$,  para evitar regimes de rotação completa do pêndulo (\textit{overshoot angular}) . Para a validação do treinamento, comparamos $\{\mathbf{x}, \dot{\mathbf{x}}, \ddot{\mathbf{x}}\}$ com $\{ \tilde{\mathbf{x}}, \tilde{\dot{\mathbf{x}}}, \tilde{\ddot{\mathbf{x}}}\}$, pois que, por mais que calculamos a diferença dos estados e as acelerações no cálculo da função custo, queremos mostrar que a rede ainda é capaz de recuperar por vezes a velocidade, já que a informação de $\dot{\mathbf{x}}$ está implícita na equação (\ref{eq:SINDy_2ndDeriv}). Assim, mesmo que não seja explicitamente parte do treinamento tomar a diferença com a primeira derivada, recuperamos um comportamento aproximado, o que demonstra que o aprendizado é robusto.

No nosso exemplo, usamos a biblioteca de polinômios até o grau 4 e a função seno. Dentre 10 inicializações testadas, 8 recuperaram o modelo correto, i.e., a equação (\ref{Eq:PenduloRaw}), as outras 2 convergiram para a solução trivial. Vale ressaltar que, em algumas inicializações, a rede recuperou o modelo da equação (\ref{Eq:PenduloSmallAngle}), que é a aproximação linear da equação original, enquanto, em outras, recuperou equações do tipo $\ddot{z} = \xi_0\sin{z} + \xi_1 z$,  eliminando sistematicamente os termos de ordem superior e atribuindo zero às funções menos relevantes, garantindo a esparsidade do modelo. 
\begin{figure} [h]

        \centering
        \includegraphics[width=1\linewidth]{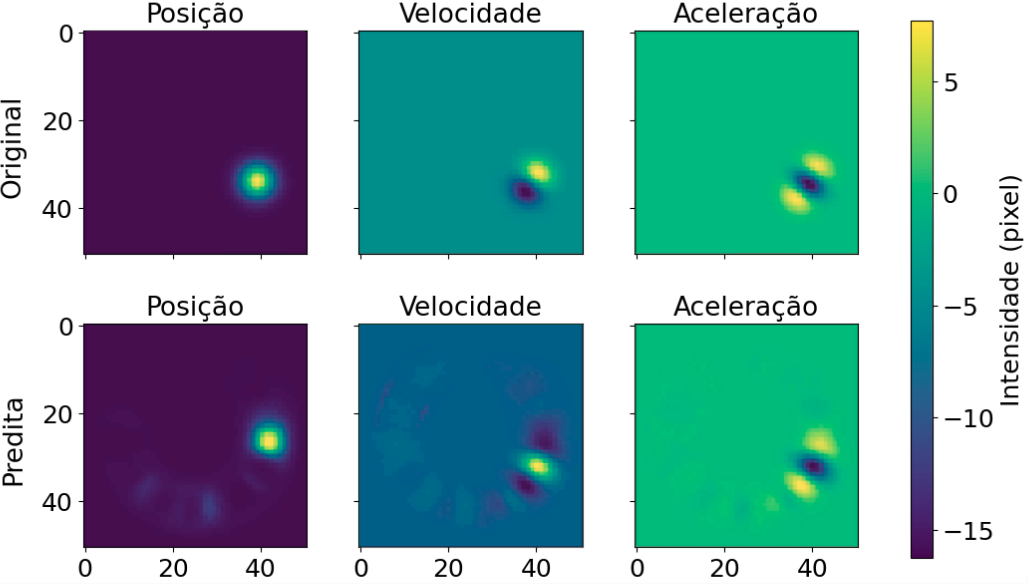}
        \caption{Imagem do Jupter notebook comparando entrada e saida de SINDy para dados de validação de posição, velocidade e aceleração, i.e., comparando $\{\mathbf{x}, \dot{\mathbf{x}}, \ddot{\mathbf{x}}\}$ com $\{ \tilde{\mathbf{x}}, \tilde{\dot{\mathbf{x}}}, \tilde{\ddot{\mathbf{x}}}\}$ para 2000 épocas. Veja que para esta otimização não foram usados os dados de velocidade explicitamente para o treino (apenas implicitamente, por meio da equação (\ref{eq:SINDy_2ndDeriv}), porém o modelo ainda é capaz de recuperá-los aproximadamente.}
        \label{fig:SINDy_ValidationData}
\end{figure}

Os resultados das imagens geradas $\{ \tilde{\mathbf{x}}, \tilde{\dot{\mathbf{x}}}, \tilde{\ddot{\mathbf{x}}}\}$ estão na Figura \ref{fig:SINDy_ValidationData}, comparados com os dados de validação $\{\mathbf{x}, \dot{\mathbf{x}}, \ddot{\mathbf{x}}\}$ em um modelo que encontrou a equação do pêndulo. 
Veja que otimizamos o modelo somente para posição e aceleração, porém, ainda assim, foi capaz de aproximar o resultado para as imagens da velocidade, o que demonstra que a rede de fato aprendeu o modelo e é capaz de reproduzir derivadas de ordem menor. Além disso, essas redes (e PINNs, em geral) requerem menos dados de treinamento do que \textit{autoencoders} tradicionais. 
Vale destacar que, neste exemplo, não utilizamos redes convolucionais (CNNs), apenas uma rede totalmente conectada, e mesmo assim foram necessárias menos épocas do que em outros casos para se obter resultados satisfatórios. Observamos que, em geral, cerca de 15\% dos dados necessários em outras arquiteturas são suficientes para obter resultados significativos, o que significa que, se 2000 épocas fossem exigidas em outra arquitetura, o SINDy deveria ser capaz de encontrar o modelo correto em um intervalo de até 300 a 500 épocas, como pode ser visto pela queda acentuada das curvas roxa e marrom  nas primeiras épocas na Figura \ref{fig:SINDy_loss}.
Este caso serve como exemplo de como o conhecimento analítico reduz a necessidade de treinamento por meio de exemplos, ou seja, uma vez que a rede aprende o modelo por otimização, ela requer menos informação para reproduzir os dados do que os modelos discutidos no restante deste artigo.

\section{Conclusões}\label{sec8}

Neste artigo, exploramos desde aspectos fundamentais a avançados das redes neurais, começando pela introdução do Perceptron até aplicações mais complexas na física. Revisamos conceitos básicos e demonstramos, por meio de exemplos didáticos, como as redes neurais podem ser empregadas em tarefas de classificação binária e regressão. Nos exemplos aplicados à física, mostramos a eficácia das redes neurais profundas aplicadas a um único sistema físico (o pêndulo simples), explorando o mesmo problema sob diversas abordagens: como encontrar o parâmetro físico de uma função a partir dos dados, como determinar a solução de uma equação diferencial e como encontrar qual a equação diferencial descreve um conjunto de dados. 

Dentre esses exemplos, destacamos a utilização das PINNs como uma ferramenta numérica particularmente promissora, embora aqui utilizada apenas para resolver equações diferenciais, mas que demonstra um leque de possibilidades ainda inexplorado. A abordagem das PINNs é particularmente relevante para físicos, ao permitir que as redes neurais aprendam de maneira mais eficiente ao integrar restrições físicas, reduzindo a necessidade de grandes conjuntos de dados e aumentando a precisão das previsões para sistemas regidos por leis conhecidas. Finalmente, vimos que, por meio de arquiteturas SINDy, redes neurais são capazes de aprender sistemas físicos até mesmo em sistemas não lineares e prover modelos analíticos, possivelmente auxiliando não só em estudos de sistemas que ainda carecem de modelos, como também provendo maior clareza sobre o produto da aprendizagem da rede, expressando-o de forma explícita em uma equação.

Deve-se notar, contudo, que o sucesso de todas estas abordagens depende criticamente da escolha de hiperparâmetros, como a profundidade e largura da rede ou a taxa de aprendizado do otimizador. Diferentemente de parâmetros físicos derivados de primeiros princípios, não existe uma fórmula geral para a seleção ótima desses valores. O processo assemelha-se mais à calibração de um aparato experimental complexo do que a uma derivação teórica. Essa realidade sublinha o papel insubstituível do cientista, cuja intuição e experiência guiam a experimentação necessária para ajustar o modelo, destacando a dimensão prática e, por vezes, artesanal, que acompanha a aplicação destas poderosas ferramentas.

Este estudo destaca a crescente importância das redes neurais como ferramentas poderosas na pesquisa em física, demonstrando seu potencial para revolucionar abordagens tradicionais em física teórica e computacional. Ao integrar métodos avançados de ML, como redes neurais, à modelagem física, este trabalho não apenas demonstra a precisão das simulações e a análise de dados complexos, mas também contribui para promover uma mudança na maneira como a ciência é conduzida. Essa integração interdisciplinar permite explorar novos horizontes em problemas que anteriormente eram intratáveis ou demandavam aproximações significativas. Assim, abre-se um caminho promissor para um novo paradigma científico, onde técnicas de IA e métodos físicos se complementam para avançar a compreensão dos fenômenos naturais. 

 \section*{Material Complementar}

 Os jupyter notebooks associados aos resultados trabalhados nos exemplos podem ser encontrados em \cite{Github}.

\section*{agradecimentos}
Agradecemos ao prof. Dr. Celso Jorge Villas Boas  pelas sugestões para a melhoria do trabalho. 

Este trabalho teve o apoio do Conselho Nacional de Desenvolvimento Científico e Tecnológico (CNPq), processos No. 465469/2014-0 e No. 311612/2021-0, e da Fundação de Amparo à Pesquisa do Estado de São Paulo (FAPESP), processos No. 2022/00209-6 e No. 2023/15739-3.



\section*{Apêndice A - Calculo da função custo}
\label{sec:Apendix:CalcFuncCusto}

Cada tipo de tarefa (como regressão, classificação, etc.) geralmente tem funções de custo mais adequadas. Aqui estão algumas das principais funções de custo usadas em ML:
\begin{table}[ht]
    \centering
    \renewcommand{\arraystretch}{2.5} 
    \begin{tabular}{|c|c|c|}
    \hline
    \textbf{Função de Ativação} & $f(x) $ & \textbf{Derivada} \\
    \hline
    Sigmoide & $ \frac{1}{1 + e^{-x}}$ & $f'(x) = f(x)(1 - f(x))$ \\
    \hline
    Tanh & $ \tanh(x)$ & $f'(x) = 1 - f(x)^2$ \\
    \hline
    ReLU & $ \max(0, x)$ & $f'(x) = \begin{cases} 1 & \text{if } x > 0 \\ 0 & \text{if } x \leq 0 \end{cases}$ \\
    \hline
    Leaky ReLU & $ \max(0.01x, x)$ & $f'(x) = \begin{cases} 1 & \text{if } x > 0 \\ 0.01 & \text{if } x \leq 0 \end{cases}$ \\
    \hline
    \end{tabular}
    \caption{Comparação das funções de ativação e suas derivadas}
    \label{tab:activation_functions}
\end{table}

Para ilustrar melhor ao leitor apresentamos os gráficos das funções da ativação e sua derivadas.
\begin{figure}[!ht]
    \centering
    \includegraphics[width=0.8\linewidth]{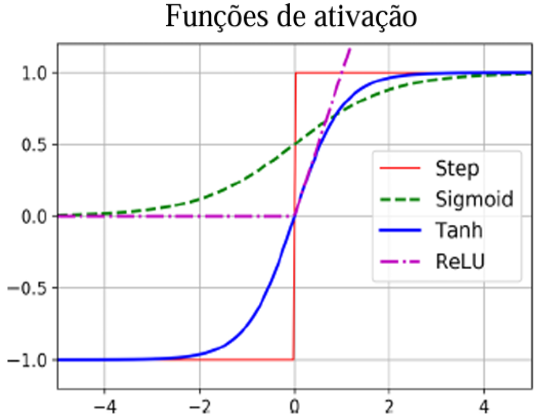}
    \includegraphics[width=0.8\linewidth]{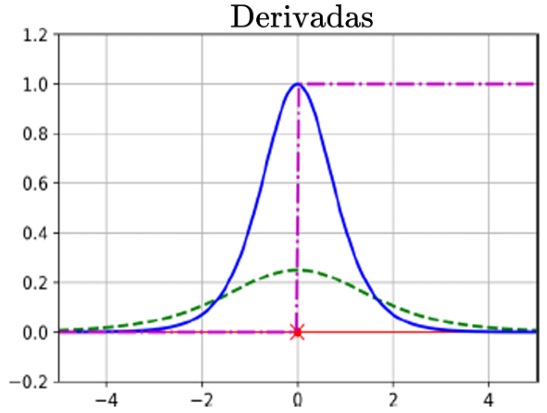}
    \caption{Funções de ativação (acima) citadas na Tabela \ref{tab:activation_functions}, e suas derivadas (abaixo) Fonte: \cite{geron}.}
\end{figure}

Podemos notar que para valores grandes algumas funções tendem a zero em suas derivadas, isso afetar o processo de treinamento. A tabela II apresenta algumas das funções custo, como Erro Quadrático Médio (Mean Squared Error - MSE) uma das funções de custo mais comuns para tarefas de regressão, ela mede a média dos quadrados das diferenças entre os valores preditos e os reais. Erro Absoluto Médio (Mean Absolute Error - MAE): Também usado em regressão, o MAE mede a média das diferenças absolutas entre os valores preditos e os reais, proporcionando uma medida de erro que é robusta a outliers. Entropia Cruzada (Cross-Entropy Loss-CCE), muito usada em problemas de classificação. Esta função de custo mede a diferença entre duas distribuições de probabilidade, a prevista pelo modelo e a distribuição verdadeira. Para a classificação binária, é conhecida como Entropia Cruzada Binária - BCE (Binary Cross-Entropy). Obs.: A Loss BCE é um caso específico da CCE, onde temos apenas duas classes. 
\begin{table}
    \centering
    \renewcommand{\arraystretch}{2.5} 
    \begin{tabular}{|c|c|c|}
    \hline
    \textbf{Loss} & \textbf{Expressão Matemática}  \\ \hline
    MSE & $\frac{1}{n} \sum_{i=1}^n (y_i - \hat{y}_i)^2$  \\ \hline
    MAE & $\frac{1}{n} \sum_{i=1}^n |y_i - \hat{y}_i|$  \\ \hline
    BCE & $-\sum [y \log(\hat{y}) + (1-y) \log(1-\hat{y})]$ \\ \hline
    CCE & $-\sum_{i=1}^n \sum_{j=1}^m y_{ij} \log(\hat{y}_{ij})$ \\\hline
    \end{tabular}
    \caption{Exemplos de funções de ativação e Funções Custo.}
\end{table}

Durante o cálculo da atualização dos pesos, precisamos do gradiente da função custo em relação aos parâmetros. Então, apresentaremos um exemplo da derivação da expressão utilizando como exemplos a função custo MSE e a função de ativação sigmoide.
\begin{align*}
    \nabla_w \mathcal{L} &=\frac{d}{dw} (y - \hat{y})^2, \\
    \nabla_w \mathcal{L} &= -2(y - \hat{y}) \frac{d}{dw}\hat{y}, \\
    \nabla_w \mathcal{L} &= -2(y - \hat{y}) \frac{d}{dw}f(w^T \mathbf{x}), \\
    \nabla_w \mathcal{L} &= -2(y - \hat{y}) f' \frac{d}{dw}(w^T \mathbf{x}), \\
    \nabla_w \mathcal{L} &= -2(y - \hat{y}) f' \mathbf{x} \frac{d}{dw}w^T, \\
    \nabla_w \mathcal{L} &= -2(y - \hat{y}) f' \mathbf{x}. 
\end{align*}
Fornecida a expressão, para os $w$ podemos fazer o mesmo para $b$.
\begin{align*}
    \nabla_b \mathcal{L} &= \frac{d}{db} (y - \hat{y})^2, \\
    \nabla_b \mathcal{L} &= -2(y - \hat{y}) \frac{d}{db}\hat{y}, \\
    \nabla_b \mathcal{L} &= -2(y - \hat{y}) \frac{d}{db}f(w^T \mathbf{x}), \\
    \nabla_b \mathcal{L} &= -2(y - \hat{y}) f' \frac{d}{db}(w^T \mathbf{x}), \\
    \nabla_b \mathcal{L} &= -2(y - \hat{y}) f' \mathbf{x} \frac{d}{db}w^T, \\
    \nabla_b \mathcal{L} &= -2(y - \hat{y}) f'. 
\end{align*}

Assim,
\begin{align*}
    \nabla_w \mathcal{L} &= -2(y - \hat{y}) f' \mathbf{x},  \\
    \nabla_b \mathcal{L} &= -2(y - \hat{y}) f' ,
\end{align*}

\noindent onde $f'$ depende de qual função de ativação está sendo usada. Na tabela \ref{tab:activation_functions} temos alguns exemplos, para ilustrar ao leitor usaremos o Sigmoide  $f'= f(1- f)$, onde $f=\hat{y}$
\begin{align*}
    \nabla_w \mathcal{L} &= -2(y - \hat{y}) \hat{y}(1- \hat{y}) \mathbf{x},  \\
    \nabla_b \mathcal{L} &= -2(y - \hat{y}) \hat{y}(1- \hat{y}).
\end{align*}

\end{document}